\documentclass[11pt]{iopart}
\usepackage{iopams}  

\usepackage{graphicx,verbatim,color}

\usepackage{dsfont}

\usepackage{epstopdf}
\usepackage{soul}

\begin{document}

\title{Probing the geometry of the Laughlin state}

\author{Sonika Johri$^{1}$, Z. Papi\'c$^{2}$, P. Schmitteckert$^{3}$, R. N. Bhatt$^{1}$, and F. D. M. Haldane$^{4}$}
\address{$^1$ Department of Electrical Engineering, Princeton University, Princeton, NJ 08544, USA} 
\address{$^2$ School of Physics and Astronomy, University of Leeds, Leeds, LS2 9JT, United Kingdom}
\address{$^3$ Institut f{\"u}r Theoretische Physik, Heinrich-Heine-Universit\"at, 40225 D\"usseldorf, Germany}
\address{$^4$ Department of Physics, Princeton University, Princeton, NJ 08544, USA} 


\date{\today}

\begin{abstract}

It has recently been pointed out that phases of matter with intrinsic topological order, like the fractional quantum Hall states, have an extra dynamical degree of freedom that corresponds to quantum geometry. Here we perform extensive numerical studies of the geometric degree of freedom for the simplest example of fractional quantum Hall states -- the filling $\nu = 1/3$ Laughlin state. We perturb the system by a smooth, spatially dependent metric deformation and measure the response of the Hall fluid, finding it to be proportional to the Gaussian curvature of the metric. Further, we generalize the concept of coherent states to formulate the bulk off-diagonal long range order for the Laughlin state, and compute the deformations of the metric in the vicinity of the edge of the system. We introduce a ``pair amplitude" operator and show that it can be used to numerically determine the intrinsic metric of the Laughlin state. These various probes are applied to several experimentally relevant settings that can expose the quantum geometry of the Laughlin state, in particular to systems with mass anisotropy and in the presence of an electric field gradient. 
\end{abstract}

\maketitle

\section{Introduction} 

Fractional quantum Hall effect (FQHE) is the phenomenon where interacting electrons form many-body liquid phases in two spatial dimensions~\cite{tsui_prl, stormer_rmp}. The best understood (and experimentally the most robust) of all such states is the $\nu=1/3$ Laughlin state~\cite{laughlin}. Phases in the FQHE display many interesting properties related to topological order~\cite{wen_book} and fractionalization, for instance, their fundamental excitations carry fractional electric charges~\cite{laughlin} and obey the fractional statistics when they are braided around one another~\cite{arovas_fractional}. The fundamental FQHE physics results from the combined effect of Coulomb interaction between electrons, and the Landau level quantization in strong magnetic fields that completely suppresses the kinetic energy of the electrons (for an introduction to FQHE physics see, e.g., Refs.~\cite{prange, qh_book, girvin_review, jainbook}). 

A traditional method that has played an important role in understanding many FQH states has been the formulation of first-quantized many-body wave functions for the ground states of the system at various filling fractions. This approach was pioneered by Laughlin \cite{laughlin} who proposed one such class of wave functions for the filling fractions $\nu=1/(2m+1)$, $m=1,2,3\ldots$. In order to write down these wave functions, it is customary to assume that a two-dimensional electron gas (2DEG) that hosts the FQHE can be viewed as a continuum system in the infinite plane. Historically, this has lead to an additional assumption that the 2DEG is also rotationally invariant around the axis perpendicular to the plane of the 2DEG. Under these assumptions, the Laughlin wave function becomes a Jastrow polynomial that involves the products of $(z_i-z_j)$, where $z$ is the complex 2D electron coordinate [see Eq. (\ref{eq:laughlin}) below for an explicit form of the wave function]. Because of rotational invariance, $z$ is fixed to be $x\pm iy$ (sign depending on the direction of the magnetic field). This type of wave functions has been microscopically very successful in modelling the exact ground state of the system computed numerically by diagonalizing the Coulomb interaction Hamiltonian. However, as variational wave functions, the Laughlin states are rather unusual because they lack any optimizing parameters. Although this surprising feature of the Laughlin states has been noted a long time ago, it is only very recently~\cite{haldane_geometry} been appreciated that the Laughlin wave functions do indeed contain a variational parameter. In order to expose this parameter, one must lift the assumption of the rotational invariance of the system. 

\begin{figure}
\centering
\includegraphics[width=0.8\linewidth]{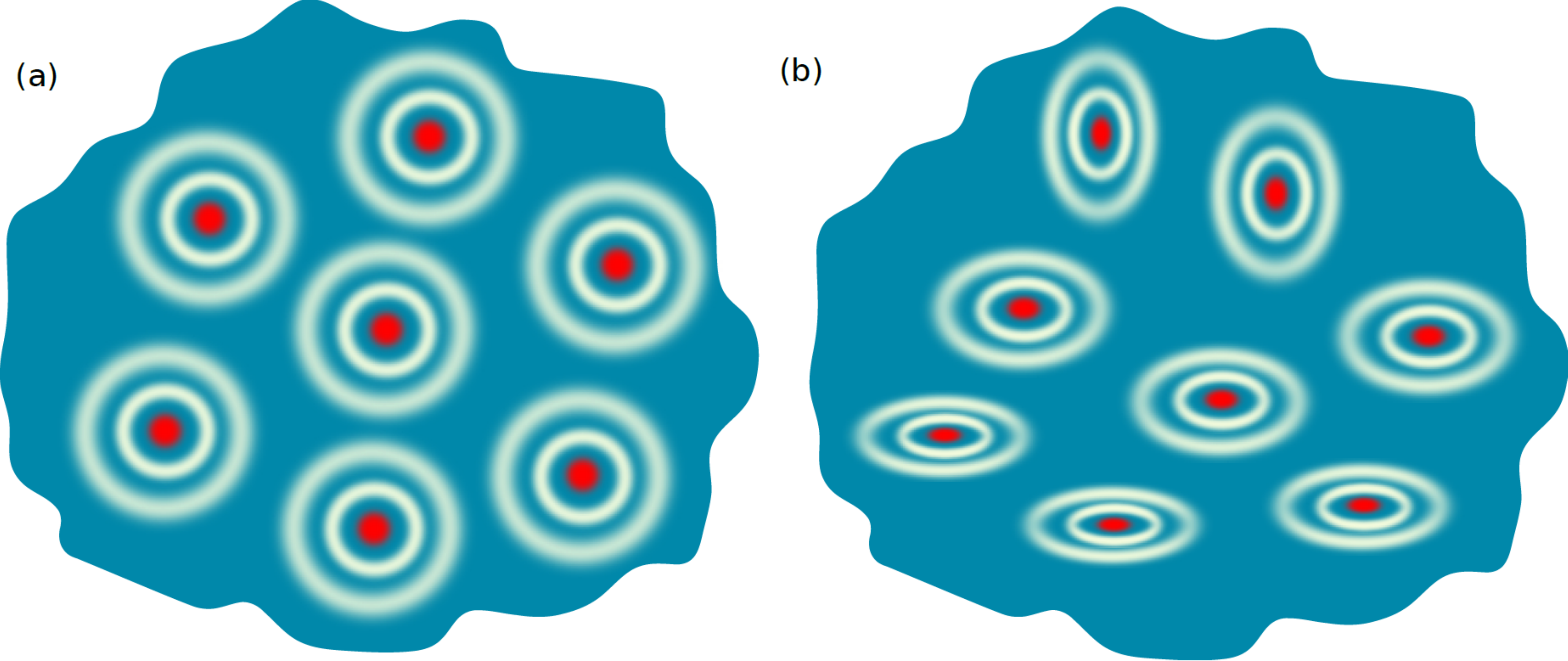}
\caption{(Color online) The cartoon picture of the Laughlin state in terms of fundamental droplets [after Ref. \cite{haldane_geometry}]. Each droplet contains one electron in the center of an area corresponding to three magnetic flux quanta. (a) The isotropic (rotationally symmetric) Laughlin state where the droplets are circles. (b) Laughlin state with a more general, spatially dependent metric where the droplets become elliptical.}
\label{fig:laughlindroplets}
\end{figure}

Because the Laughlin 1/3 state describes a gapped topological phase, i.e., a liquid which is robust to any local perturbations that do not close the gap of the system, its existence is not sensitive to the geometric details of the system. The basic physics of the Laughlin state is that it efficiently places electrons apart from one another, yet avoids breaking the translational symmetry in doing so. According to Ref. \cite{haldane_geometry}, we can thus characterize the Laughlin state in terms of fundamental droplets which are schematically illustrated in Fig.~\ref{fig:laughlindroplets}. Each droplet contains one electron in an area corresponding to 3 magnetic flux quanta (in general, for more complicated states at fillings $\nu=p/q$, a droplet would contain $q$ orbitals inhabited by $p$ electrons). When the system is rotationally isotropic, the shape of the droplet is circular. However, in a more general setting [Fig.~\ref{fig:laughlindroplets}(b)] the metric may be spatially dependent. For example, the 2DEG may be wrapped around a curved surface, the direction of the magnetic field may be slightly tilted at various points in space, etc. In these circumstances, it is natural to expect that the shape of the fundamental droplets will also vary depending on the location of their center. Therefore, the shape of the droplets is a hidden variational parameter that characterizes the Laughlin state and this parameter can be tuned to yield the best variational description of the actual ground state of the system.  

Our cartoon of the fundamental droplets of the Laughlin state can be formalized into a phenomenological picture of the FQHE as a fluid of particle-flux composites with finite area. These composite particles are known as ``composite bosons"~\cite{zhk} or ``composite fermions"~\cite{jain89}, depending on how many orbitals surround an electron. In general, the composites carry information about numbers $p$ and $q$ that define the filling fraction $\nu=p/q$. Moreover, the occupancy pattern inside the droplet is also connected to the concept of ``topological spin" (which is also related to the ``shift"~\cite{wen_zee} when the quantum Hall state is placed on a sphere). The finite area of the droplet means that a metric [that of the guiding center defined below] is required to specify the shape. In addition to the shape, the guiding center metric is also related to the spin. The deformation of the shape of the composite particle couples to spin, thus connecting topology and geometry. The fluctuations of the quantum metric and its coupling to spin gives rise to dynamics in the FQHE phases.

More precisely, in the full problem of a 2DEG in the magnetic field, one encounters, in fact, two distinct metrics. This can be seen as follows. The phase space for each particle confined to two dimensions in a transverse magnetic field consists of two sets of non-commuting real-space coordinates. The position of an electron can be separated into the cyclotron and the guiding center coordinates respectively as $r^a=R^a+\tilde{R}^a$, $a=x,y$. The cyclotron coordinates are related to the canonical momenta by $\tilde{R}^a=\frac{\ell_B^2}{\hbar} \epsilon^{ab} \pi_b$, with $\epsilon$ being the two-dimensional Levi-Civita symbol, and $\ell_B=\sqrt{\hbar/eB}$ is the magnetic length. The commutation relations $[R^a,R^b]=-i\ell_B^2\epsilon^{ab}$ and $[\tilde{R}^a,\tilde{R}^b]=i\ell_B^2\epsilon^{ab}$ hold. These are independent Hilbert spaces, in each of which the real-space coordinates do not commute with each other, leading to quantum fluctuations of the metric of each. The cyclotron coordinates are present in the term for kinetic energy. On the other hand, the guiding center coordinates are present in the interaction term. 

One way to characterize the difference between integer quantum Hall effect (IQHE)~\cite{klitzing_prl, klitzing_rmp} and FQHE is by the part of the Hilbert space in which the relevant dynamics takes place. For IQHE, the guiding center degrees of freedom are frozen because the LLs are fully filled, and the dynamics is governed by cyclotron coordinates. For FQHE, it is the opposite: the cyclotron degree of freedom is frozen by the strong magnetic field, and the dynamics is governed by the guiding center coordinates. The fluctuations of the guiding center metric give rise to phenomena such as the FQHE bulk neutral excitations in long wavelength limit \cite{bo_thesis}. The guiding-center metric is also related to the Hall viscosity \cite{avron, tokatly, read_viscosity, haldane_viscosity, readrezayi_viscosity, yeje}.

Recently, anisotropy and geometry in FQHE systems have received much attention. Several papers have studied FQHE systems in curved spaces \cite{son, gromov1, bradlyn, gromov2}. Transport coefficients have been calculated as a response to variations of spatial geometry \cite{wiegmann, wiegmann2, bradlyn2}. Anisotropic metric has been connected to the band structure of materials giving rise to the FQHE \cite{boyang, haowang, apalkov, areg}, the tilting of the magnetic field \cite{papic_tilted} and the so-called nematic phases in higher Landau levels \cite{maciejko}. Anisotropic mass has been shown to affect the shape of the composite fermion surface \cite{kunyang1}. In experiment, the anisotropy of the Fermi contour has been studied in GaAs quantum wells using an L-shaped Hall bar and periodic strain engineered using a grating of electron-beam resist \cite{shayegan, shayegan2, shayegan3}. Finally, there are recent proposals for experimental implementations of time-dependent external metric via acoustic crystalline waves~\cite{kunyang2}. 

In this paper, we study the geometric degree of freedom that characterizes the shape of the fundamental droplets of the Laughlin $\nu=1/3$ state. We design several numerical experiments and probes that can be used to detect the fluctuations of this degree of freedom of FQHE states. For most of the calculations, we use the geometry of an open cylinder where the metric perturbation can be conveniently introduced, but our results are sufficiently general that they apply to any geometry. One advantage of the cylinder geometry is that all the single-particle orbitals in momentum-space have the same shape, unlike the disk or the sphere. Therefore, spatially-varying properties can be studied without interference from other geometrical effects. In Section \ref{sec:cyl}, we provide a self-contained review of the quantum Hall problem in the cylinder geometry. In Section \ref{sec:coh} we introduce the generalization of coherent states in the spirit of the fundamental droplets sketched in Fig.~\ref{fig:laughlindroplets}. We show that such generalized coherent states can be used to formulate an ``off-diagonal long range order" (ODLRO) parameter that is quantized at long distances for the Laughlin state at $\nu=1/3$, but vanishes for a compressible state at the same filling factor. We also study the squeezing of coherent states in the vicinity of an edge of the cylinder. In Section \ref{sec:response} we spatially perturb the metric in a controlled and smooth way, and study the response of the FQH fluid. In agreement with analytical expectations, we find the response to be proportional to the Gaussian curvature of the perturbed metric. In Section ~\ref{sec:pair} we introduce a different operator -- the ``pair amplitude" operator -- and use it to measure the intrinsic geometry fluctuations of the Laughlin state when the mass tensor is anisotropic or the system is perturbed by an electric field gradient. Our conclusions are summarized in Section \ref{sec:conc}.

\section{Interacting electrons on a cylinder}\label{sec:cyl}

We consider a 2DEG confined to the surface of a finite cylinder with radial magnetic field. We set the $x$-axis to be along the axis of the cylinder and $y$-axis to be the periodic direction. This is equivalent to working in the Landau gauge with vector potential $\mathbf{A}=Bx\hat{y}$. The momentum along the $y$-direction, $k_y$, is a good quantum number. The single particle wave functions (``orbitals") have the form~\cite{qh_book}
\begin{equation}\label{eq:single}
\phi_{n,k_y}=\frac{1}{\sqrt{L\ell_B \sqrt{\pi}}}e^{ik_yy-\frac{1}{2}\left(\frac{x}{\ell_B}-\ell_B k_y\right)^2} H_n\left( \frac{x}{\ell_B}-\ell_B k_y \right),
\end{equation}
where $l_B=\sqrt{\hbar/(eB)}$ is the magnetic length, $H_n$ is the Hermite polynomial and $n=0,1,2...$ labels the Landau levels. In the following, we restrict to the lowest Landau level (LLL) corresponding to $n=0$. The allowed values of $k_y$ are $2\pi m/L$, where $L$ is the circumference of the cylinder and $m$ is an integer that resolves the degeneracy within the LLL. 
We use a finite number of orbitals $N_{\rm orb}$, which implies that the size of the system along the $x$-direction is approximately $H=(2\pi/L)N_{orb}$. In the LLL, the wave function $\phi_{k_y}$ is then the product of a Gaussian along the $x$-axis localized at $x=(2\pi ml_B^2/L)$  and a plane wave along the $y$-axis. The value of $N_{\rm orb}$ is set by the filling factor but also by the nature of the given state, due to the topological quantum number known as the ``shift"~\cite{wen_zee}. For example, for the $\nu=1/3$ Laughlin state, $N_{\rm orb}=3N-2$ gives the correct ground state. Note that the shift can be determined from the occupancy pattern of the fundamental droplet, which according to Fig.~\ref{fig:laughlindroplets} is $100100...1001001$ for the Laughlin state.

The two relevant length scales at this stage are the magnetic length $\ell_B$, which sets the width of the wave function, and $L$, the circumference of the cylinder which controls the distance and hence the overlap between the single-particle wave functions. Henceforth, we set $\ell_B=1$. We are interested in solving for the ground state and possibly a few low-lying excited states of a system of $N$ interacting electrons. This must be done numerically, and we resort to two techniques: exact diagonalization and density matrix renormalization group (DMRG). Exact diagonalization is an unbiased method of finding the eigenstates of the many-body Hamiltonian, but limited to small systems because of the exponential increase in size of the Hilbert space. DMRG~\cite{dmrg_white} is a variational optimization over a class of ``weakly-entangled" states known as ``matrix product states"~\cite{mps1,mps2,mps3}. In the past, DMRG has been applied to FQH systems in various geometries~\cite{dmrg_shibata, dmrg_feiguin, dmrg_kovrizhin, dmrg_zhao}, but the convergence of the method was found to be the best for the cylinder geometry  \cite{dmrg_cyl, fqhe_mps1, fqhe_mps2, dmrg_zaletel}.

A useful point of departure for the study of FQHE phases are the ground states of model Hamiltonians that have high overlap with the ground state of the Coulomb interaction. For example, at $\nu=1/3$, it can be shown that the ground state for $N=6,7,\ldots, 12$ electrons interacting via Coulomb repulsion has $\gtrsim 95\%$ overlap with the ground state of the following Hamiltonian:
\begin{equation}\label{eq:ham_laughlin}
H_L= \sum_{i<j} \nabla^2 \delta(\mathbf{r}_i-\mathbf{r}_j).
\end{equation}
Note that this Hamiltonian is singular; however, once its matrix elements are evaluated between the single-particle wave functions in Eq. (\ref{eq:single}), all divergences are automatically removed. The Fourier transform of $H_L$, which is better known as the $V_1$ Haldane pseudopotential \cite{haldane_sphere}, is given by the first Laguerre polynomial.

The ground state of the Hamiltonian in Eq. (\ref{eq:ham_laughlin}) can be analytically shown to have exactly zero energy \cite{haldane_sphere}. In the infinite plane, the wave function of this ground state is the Laughlin state,
\begin{equation}\label{eq:laughlin}
\Psi_L (z_1,\ldots, z_N) =\prod_{i<j}^N (z_i-z_j)^3 e^{-\sum_k |z_k|^2/4\ell_B^2},
\end{equation}
where $z_j$ denote complex coordinates of electrons in the 2D plane. Traditionally, one would write $z_j = x_j + i y_j$, but note that this parametrization is only valid for a rotationally invariant system such that the contour of constant $|z|^2$ is the circle. More generally, $z_j = \alpha x_j + i y_j/\alpha$ is also a valid choice, which for $\alpha\neq 1$ gives an elliptical contour of $|z|^2$. The wave function $\Psi_L$ is believed to represent all fundamental aspects of the physics of an actual $\nu=1/3$ state, even though the wave function of the physical system is far more complicated than Eq. (\ref{eq:laughlin}). The reason is that $\Psi_L$ can be adiabatically connected to the ground state of the system with Coulomb interaction without closing the excitation gap \cite{rh85}. 

Many finite-size studies focus only on model interactions such as the one above, which is reasonable since these ground states typically have high overlap with the physical ground state and in particular share the same topological properties. However, in order to account for all the details of the physical ground state, one needs to directly model the Coulomb interaction and its ground state separately. On a cylinder, the Coulomb interaction has to be implemented with care because of the infinities that arise from its long range nature. We next discuss this point in some detail.

\begin{figure}[htb]
\centering
\includegraphics[width=4 in]{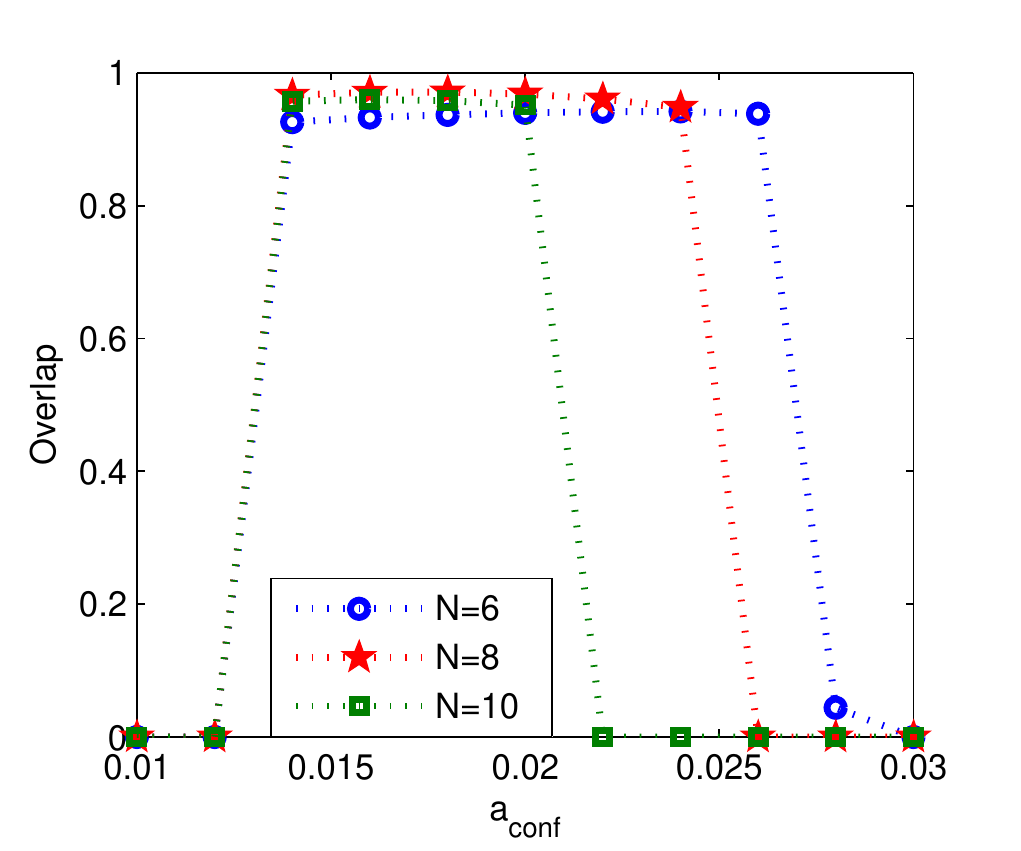}
\caption{(Color online) Overlap, $|\langle\Psi_L|\Psi_C\rangle|^2$, between the Laughlin wave function and the ground state of Coulomb interaction as a function of confinement strength for $N=6$, $8$ and $10$ electrons in $N_{\rm orb}=3N-2$ orbitals.}
\label{fig:overlaps}
\end{figure}

The Fourier transform of the rotationally-invariant Coulomb interaction in two-dimensions is $\tilde{V}(k)=\frac{2 \pi}{k}$, where $k=\sqrt{k_x^2+k_y^2}$. Therefore, it can be written as
\begin{equation}\label{eq:Vr}
V(r)=\frac{1}{(2\pi)^2}\int d^2\mathbf{k} \; \frac{2 \pi}{k}  e^{i \mathbf{k}. \mathbf{r}}
\end{equation}
The integral in Eq. \ref{eq:Vr} diverges at $k_x,k_y=0$. Since there is no periodicity along the $x$-axis, the $k_x$ integral cannot be converted into a sum, in which case the singularity could be removed simply by ignoring the problematic term in the sum which has both $k_x=0$ and $k_y=0$. Instead, we use the following strategy that works in the continuum limit.

In the second quantized notation, the Hamiltonian for the Coulomb interaction reads
\begin{equation}\label{eq:second}
H=\sum_{m_1,m_2,m_3,m_4}' V_{m_1,m_2,m_3,m_4}c^{\dagger}_{m_1}c^{\dagger}_{m_2}c_{m_4}c_{m_3}.
\end{equation}
The matrix element is given by 
\begin{eqnarray}
V_{m_1,m_2,m_3,m_4} &=& \int d \mathbf{r}_1 d \mathbf{r}_2 \phi^{*}_{m_1}(\mathbf{r}_1) \phi^{*}_{m_2}(\mathbf{r}_2) V(\mathbf{r}_1-\mathbf{r}_2) \phi_{m_3}(\mathbf{r}_1) \phi_{m_4}(\mathbf{r}_2) \nonumber\\
&=& \frac{e^{-\beta^2/2}}{2\pi L}\int_{0}^{\infty}dk_x \frac{e^{-k_x^2/2} \cos(k_x \gamma)}{\sqrt{k_x^2+\beta^2}}
\end{eqnarray}\label{Vall}
where $\beta=\frac{2\pi}{L}(m_3-m_1)$ and $\gamma=\frac{2\pi}{L}(m_3-m_2)$. The prime on the sum in Eq. (\ref{eq:second}) denotes the conservation of momentum $m_1+m_2=m_3+m_4$, resulting from the integration over $k_y$.

The singularity appears when both $\beta$ and $k_x$ become zero. At $\beta=0$, the interaction can be separated into a singular and non-singular part:
\begin{eqnarray}
V_{m_1=m_3,m_2=m_4} &=&\frac{1}{2\pi L}\left(\int_{0}^{\infty}{dk_x \frac{1}{k_x}}+ \int_{0}^{\infty}{dk_x \frac{e^{-k_x^2/2} \cos(k_x \gamma)-1}{k_x}}\right).
\end{eqnarray}
Since the singular part is independent of all $m_i$, we can discard it without affecting the eigenstates of the Coulomb Hamiltonian. However, doing this means that we are subtracting an arbitrary term from the energies of the eigenstates. Therefore, the ground state energy cannot be compared to that obtained from other geometries.

Another subtlety is that in order to model long-range interactions in an open system, we need a confining potential. This is because long-range interactions have a tendency to pile up the electrons near the edges, which can easily destroy the liquid ground state like the Laughlin state. For the model interaction $H_L$ which is of ultrashort range, no confinement is required. However, long range interactions such as the Coulomb do require a confining potential along the axis of the cylinder. In the absence of such confinement, electrons would lower their energy by moving symmetrically to the edges of the system, leading to reconstructed edges \cite{edge, zixiang_edge}, similar to the behavior on the disk geometry. 

We compared several confinement schemes in detail, including a simple parabolic confinement, Hartree-type confinement (in which the finite cylinder is considered to be embedded in an infinite one and the external uniform charge provides an effective confinement), and a co-axial sheet of charge. From these, we chose a simple parabolic one:
\begin{equation}\label{eq:conf}
V=a_{\rm conf} (x/\ell_B)^2
\end{equation}
which is the easiest to implement, yet sufficient to stabilize the FQH state depending on the value of $a_{\rm conf}$. Typically, finding the appropriate $a_{\rm conf}$ to get the correct ground state will require some fine tuning. One way to check that the Coulomb ground state, $\Psi_C$, at $\nu=1/3$ is in the right phase is to calculate its overlap with the Laughlin wave function, $|\langle\Psi_L|\Psi_C\rangle|^2$. Using the simple parabolic potential, we can obtain overlaps $>90\%$ shown in Fig. \ref{fig:overlaps}. The overlap is high only in the narrow range of the confining potential, indicating that the confinement has to be finely tuned to obtain the desired FQHE ground state.

\begin{figure}
\centering
\includegraphics[width=4in]{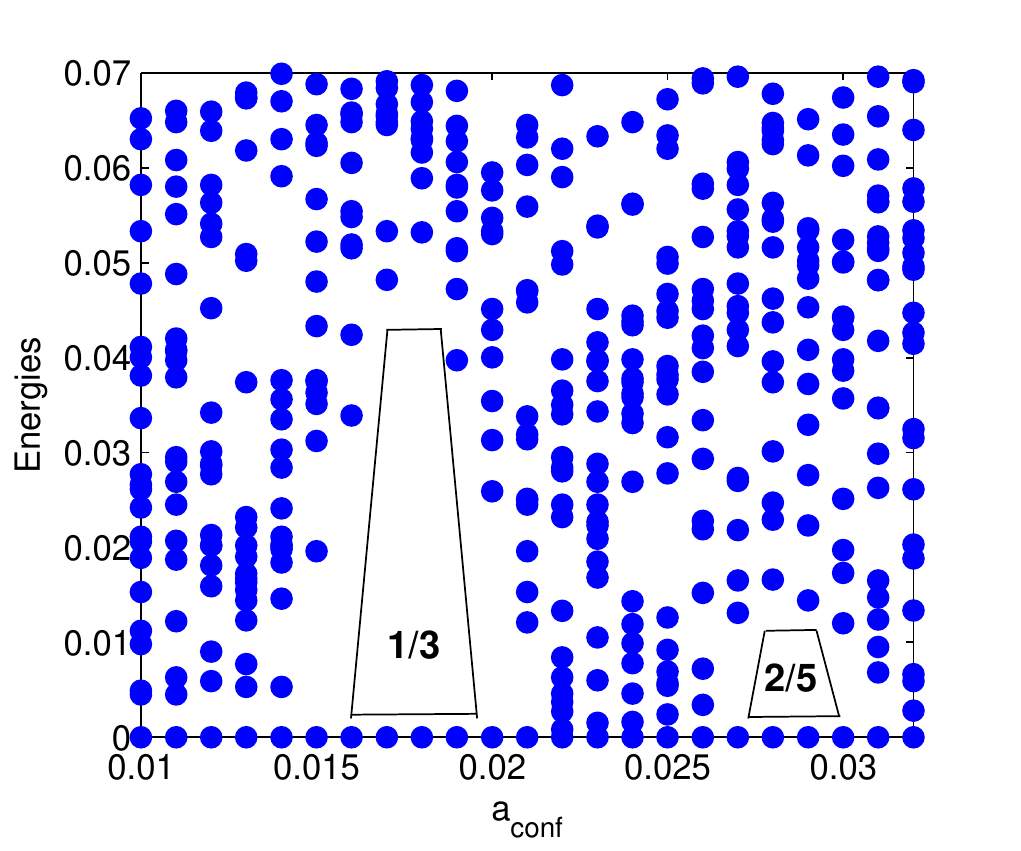}
\caption{(Color online) Transitions between $1/3$ and $2/5$ hierarchy states as a function of confinement. Low-lying energy spectrum for $N=10$ electrons in $N_{\rm orb}=28$ orbitals for the Coulomb interaction is shown as a function of parabolic confinement strength. Each of the blue dots represents an energy level of the system at that value of the confinement.}
\label{fig:conf_transition}
\end{figure}
The confining potential is also a convenient way to induce topological phase transitions between FQHE hierarchy states such as $1/3\rightarrow 2/5\rightarrow 3/7$, etc. Imagine that we are in the thermodynamic limit when the cylinder is very long, but the density of electrons is just right to be in the  $\nu=1/3$ ground state. Once we start increasing the magnitude of the confining potential, the $\nu=1/3$ state will display some rigidity to the perturbation because it is an incompressible fluid with a finite gap for all excitations. For sufficiently strong confinement, the ground state will become too squeezed and can no longer support the incompressible state. At this point, the state is gapless. Upon even further squeezing, the system will make a transition to a new gapped state which can be viewed as the condensate of the quasiparticles of the 1/3 state, i.e. the $\nu=2/5$ hierarchy state. The scenario then repeats until we reach $\nu=3/7$ state, etc. 

\begin{figure}
\centering
\includegraphics[width=5 in]{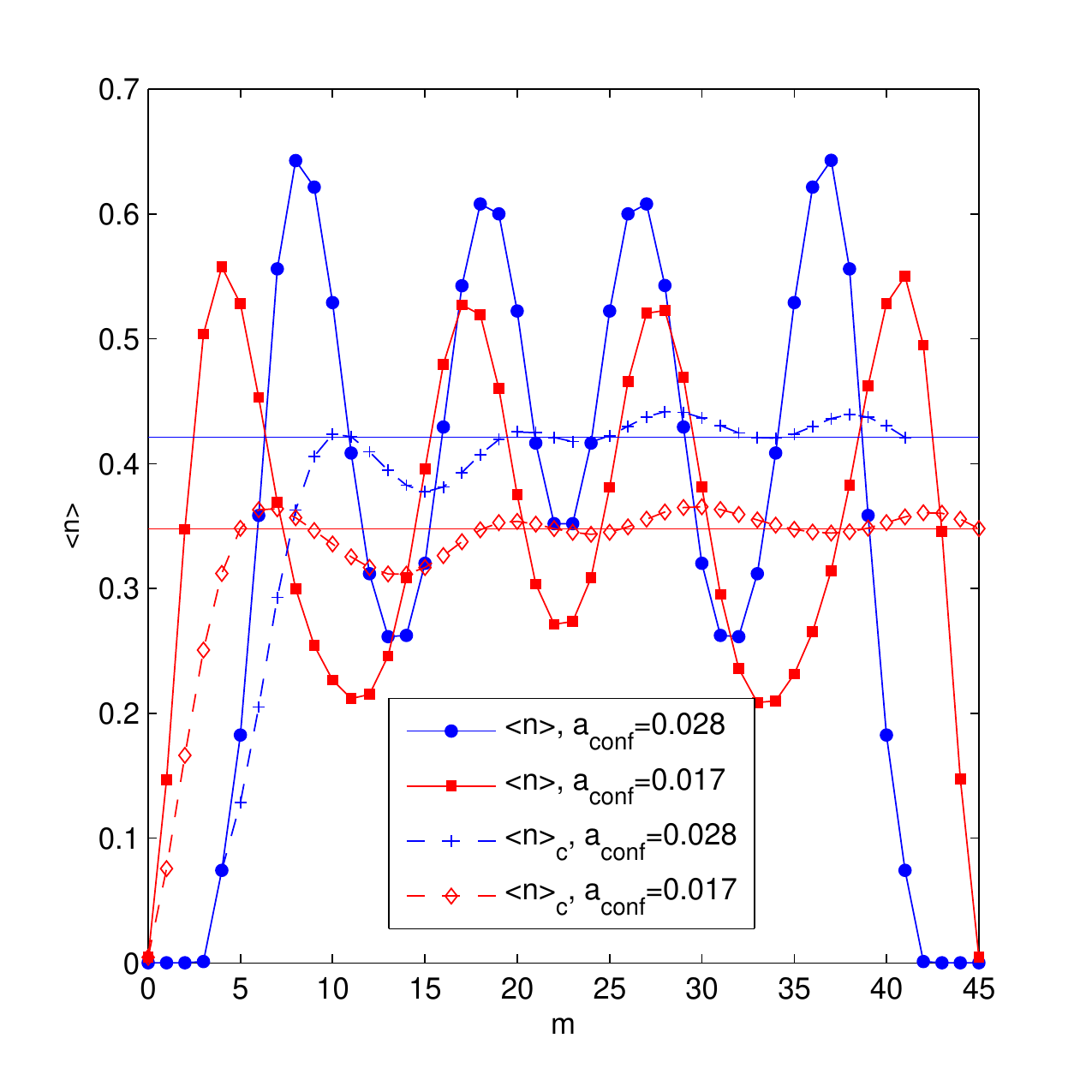}
\caption{(Color online) Average occupation numbers $\langle n_m \rangle$ ($m=0,1,\ldots,N_{\rm orb}-1$) for $N=16$ electrons in $N_{\rm orb}=46$ orbitals for Coulomb interaction with different confinement strengths obtained using DMRG (maximum number of states kept is 2500, discarded entropy is $10^{-12}$). Also shown is the cumulative average of the occupation numbers, $\langle n_m \rangle_c$ which oscillates about the expected filling factor. Note that the actual filling for the finite size system is slightly higher than that in the infinite limit because of the shift \cite{wen_zee}.}
\label{fig:density_both}
\end{figure}
In a finite system, we can resolve the first transition between the members of the hierarchy states. Fig. \ref{fig:conf_transition} shows the energy spectrum for 10 electrons as a function of the confinement parameter $a_{\rm conf}$ in Eq. (\ref{eq:conf}). The system contains $N_{\rm orb}=28$ orbitals, which is enough to realize the Laughlin state for $N=10$ electrons. As the confinement increases, the electrons are effectively restricted to a smaller number of orbitals in the middle of the cylinder, thus effectively increasing the filling factor. Gaps between the ground state energy and the rest of the spectrum open and close at different confinement strengths. Gaps are present at $a_{\rm conf}=0.017$ and $a_{\rm conf}=0.028$. These correspond to the FQHE states at $1/3$ and $2/5$. 

We can confirm the correct nature of the ground states, e.g., by looking at the occupation numbers in the bulk for the same value of the confinement but in a much larger system that can be studied with DMRG. The results for $N=16$ electrons in 46 orbitals are shown in Fig. \ref{fig:density_both}. This plot shows that the number of orbitals with non-zero occupation result in the right fractional fillings, and the occupation numbers in the bulk indeed oscillate about the correct fractional values, $1/3$ and $2/5$. As the final outcome, for very large confinement, the system will ultimately transition to the $\nu=1$  quantum Hall state in the middle of the cylinder. Larger systems would naturally have the ability to display more gapped fractional states as a function of confinement before this final integer quantum Hall state is achieved.

Apart from the confinement potential, the circumference of the cylinder $L$ is another tunable parameter. Its effect is two-fold: it modifies the matrix elements of the interaction potential, and also determines to what extent the edge effects penetrate into the bulk of the system. These effects are systematically studied in Appendix A.


\section{Coherent states and off-diagonal long-range order}\label{sec:coh}

In this section, we introduce a generalization of coherent states which are used to formulate the off-diagonal long-range order \cite{yang} (ODLRO) of the $\nu=1/3$ Laughlin state. The ODLRO is normalized in such a way that it approaches the value $1/3$ for large distances if the system is the Laughlin state, while it decays to zero in a compressible phase at the same filling factor. This definition of the ODLRO naturally applies to the case when the system is anisotropic. We also show that it can be used as an indicator of local changes in the shape of fundamental droplets close to the edge of an open cylinder.

The kinetic energy, corresponding to the cyclotron degree of freedom, for a single electron is given by 
\begin{equation}
K=\frac{1}{2m_e}g^{ab}\pi_a\pi_b
\end{equation}
where $g^{ab}$ is the cyclotron metric and $\pi_a$ are the canonical momenta. At high magnetic fields and for $\nu<1$, this cyclotron degree of freedom is frozen out and $K$ is a constant when electrons are confined to the LLL. The remaining degrees of freedom are the guiding center operators $R^a=r^a-\frac{\ell_B^2}{\hbar}\epsilon^{ab}\pi_{b}$, where $\epsilon^{ab}$ is the antisymmetric tensor of rank 2. Using the guiding center coordinates, we can define the raising and lowering (harmonic oscillator) operators $b$ and $b^{\dagger}$ in the usual way~\cite{jainbook}.

\begin{figure}
\hspace{-1 in}
\includegraphics[width=7.5 in]{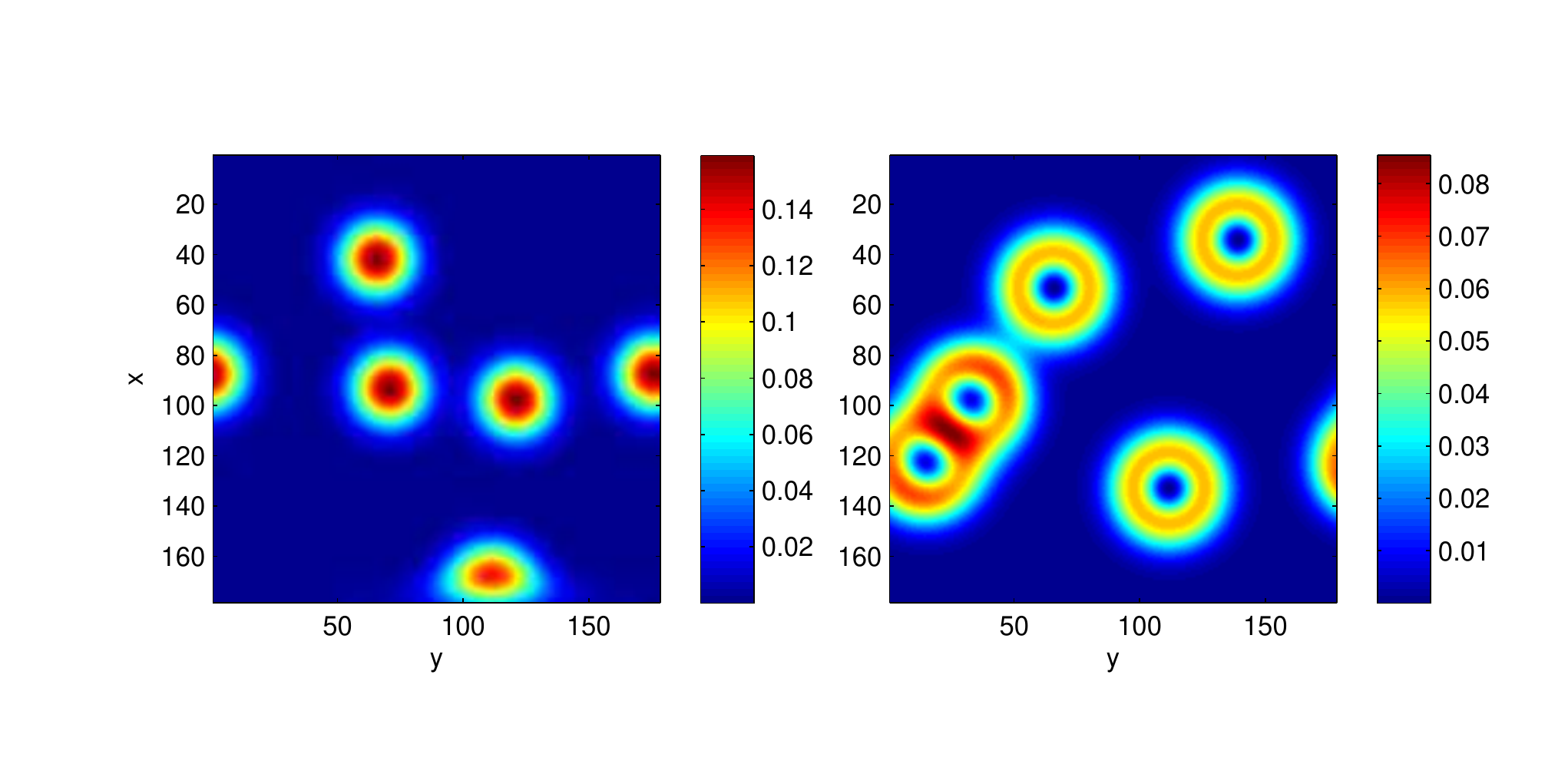}
\vspace{-0.8 in}
\caption{(Color online) Density plot of coherent states with $m=0$ (left panel) and $m=1$ (right panel) randomly scattered across the cylinder surface. }
\label{fig:droplets}
\end{figure}
The coherent state $\phi_\rho(z)$ centered at a position $\rho=x-iy$ is an eigenstate of the lowering operator \cite{girvin_review}:
\begin{eqnarray}
\phi_{\rho}(z)&=\frac{1}{\sqrt{2\pi}}\exp{\left[\frac{1}{2}\bar{\rho}z-\frac{1}{4}|z|^2-\frac{1}{4}|\rho|^2\right]}\\
b\phi_{\rho}(z)&=\frac{\bar{\rho}}{\sqrt{2}}\phi_{\rho}(z)
\end{eqnarray} 

This coherent state can also be expressed as the lowest Landau level projection of the Dirac delta function \cite{girvin_review}. 
Several coherent states, located at various points on the surface of the cylinder, are depicted in Fig. \ref{fig:droplets} (left). In the bulk, the states have circular symmetry and their intensity peaks at their centers. They get squeezed as they approach the $x$-edge, and wrap-around the periodic $y$-edge.

We can construct a more general family of coherent states parametrized by an integer $m$:
\begin{equation}\label{eq:coh}
|m,\rho\rangle = e^{i \vec{\rho} \vec{R}} (b^{\dagger})^m |0\rangle=(b^{\dagger}-\bar{\rho})^m |0\rangle,
\end{equation}
where $\vec{R}$ is the guiding-center coordinate vector.
These states are still centered about the point $\rho$, but their intensity is spread over the concentric circle around $\rho$. The radius of the circle is fixed by $m$, and scales as $\sim \sqrt{m}\ell_B$.

The motivation for introducing the objects in Eq. (\ref{eq:coh}) is that they will provide a mathematical description of our cartoon of the fundamental droplets in Fig.~\ref{fig:laughlindroplets}. As we mentioned earlier, 
the phenomenological picture of FQH states is that of a fluid being composed of droplets -- the composites of particles and empty magnetic orbitals around them (i.e., their correlation holes). At filling factor $\nu=p/q$, each droplet consists of $p$ filled coherent states, and $q-p$ empty ones. The occupation pattern inside the droplet determines the guiding-center ``spin" $s=\frac{p}{2}(p-q)$ \cite{haldane_geometry}. A change in the occupation of the droplet requires a finite amount of energy which leads to the incompressibility of the FQHE state. Therefore, the essence of incompressibility is the finite expectation value for destroying the droplet at the origin and creating it at some point far away, similar in spirit to the conventional ODLRO~\cite{yang}. The first discussion of ODLRO for the quantum Hall effect can be found in Refs.~\cite{girvin_odlro, read_odlro}, while the first numerical studies of ODLRO in FQH states were performed in Ref. \cite{rh88} and Ref. \cite{chak_odlro}. There it was shown that ODLRO exists for the filling fractions $\nu=1/3$ and $2/5$, and disappears when the incompressibility of the state is destroyed. 

To mathematically define our generalized ODLRO, we need to introduce a projector onto a coherent state as follows:
\begin{eqnarray}
\hat{n}_m(\vec{\rho}) = |m, \rho\rangle \langle m, \rho|.
\end{eqnarray}
This projector can be more conveniently evaluated by taking the Fourier transform,
\begin{eqnarray}\label{eq:nm}
\hat{n}_m(\vec{\rho}) = \int \frac{d^2 q}{(2\pi)^2} L_m\left( \frac{q^2}{2} \right) e^{-q^2/4} e^{-i \vec{q}. \vec{\rho}} e^{i \vec{q}. \vec{R}},
\end{eqnarray}
where $L_m$ is the $m^{th}$ Laguerre polynomial. Note that $\hat{n}_m(\vec{\rho})$ defined in the previous equation is indeed an operator because it contains the exponential of the guiding center coordinate, therefore it can be expressed in terms of the magnetic translation operators. By explicit diagonalization of  $\hat{n}_m(\vec{\rho})$  we can verify that it has a single non-zero eigenvalue (that can be normalized to one), and another $N_{\rm orb}-1$ eigenvalues that are exactly zero (to numerical precision). This means that $\hat n_m$ indeed acts as a valid projection operator. Note that we can only fix the normalization in the bulk of the system (in the vicinity of the edge the eigenvalue will deviate from one in a non-universal manner). 

Using the projector $\hat n_m$ we can define an ODLRO, $C(\mathbf{\rho})$, for the $\nu=1/3$ state:
\begin{eqnarray}\label{eq:odlro}
C(\mathbf{\rho}) \equiv \langle \Psi(\mathbf{\rho})|\Psi(0)\rangle, \\
| \Psi(\mathbf{\rho}) \rangle \equiv \left(\mathds{1}-\hat{n}_2(\mathbf{\rho})\right) \left(\mathds{1}-\hat{n}_1(\mathbf{\rho})\right) \hat{n}_0 (\mathbf{\rho}) |\Psi_0\rangle,
\end{eqnarray}
where $\Psi_0$ denotes the ground state of the system for which the ODLRO is computed. The meaning of this definition is as follows. When $\hat n_0$ acts on the ground state, $\hat{n}_0 (\mathbf{\rho})\Psi_0$, it creates a particle in the $m=0$ coherent state centered at $\rho$. Next, we act on this state with $(\mathds{1}-\hat{n}_1(\mathbf{\rho}))$. Since $\hat n_1$ is a projection operator with eigenvalue 1, $(\mathds{1}-\hat{n}_1(\mathbf{\rho}))$ removes an electron from the $m=1$ state centered around the same $\rho$. [And similarly for $(\mathds{1}-\hat{n}_2(\mathbf{\rho}))$]. Therefore, the resulting $\Psi(\mathbf{\rho})$ is the ground state of the system where one electron is pinned at $\rho$, but there are no other electrons in states $m=1,2$ around that point. Thus, $\Psi(\mathbf{\rho})$ is the state with a fundamental droplet created at position $\rho$. The correlator $C(\mathbf{\rho})$ then expresses the amplitude to create such a droplet at the origin and remove it from another point $\rho$. 

\begin{figure}
\centering
\includegraphics[width=4 in]{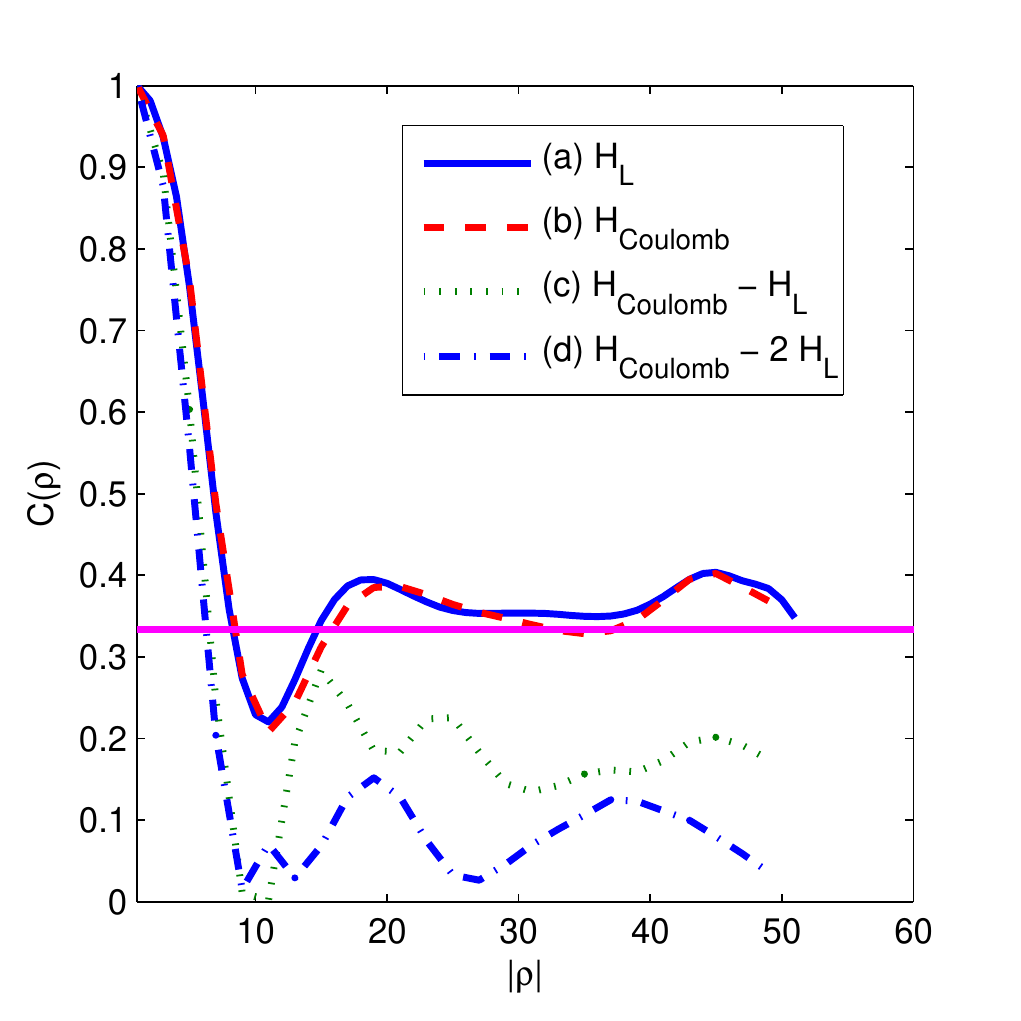}
\caption{(Color online) The ODLRO for $N_e=9$ electrons in $N_{\rm orb}=25$ orbitals for the ground states of different interactions. $H_{\rm Coulomb}-\lambda H_L$, $\lambda=1,2$,  refers to the Coulomb interaction from which some amount of short-range $H_L$ has been subtracted. The horizontal line marks $1/3$ on the y-axis. }
\label{fig:odlro_n_9}
\end{figure}
We have computed the ODLRO according to Eq. (\ref{eq:odlro}) as a function of position in Fig. \ref{fig:odlro_n_9}. We consider $N_e=9$ electrons in $N_{\rm orb}=25$ orbitals on the cylinder, and several types of interaction potentials. The Laughlin state in this figure is obtained as the ground state of the $H_L$ Hamiltonian. The Coulomb ground state is obtained for a tuned value of the confinement $a_{\rm conf}$ where it has a large overlap with the Laughlin state. In these two cases, the ODLRO is trivially 1 (in our normalization) when $\rho=0$, but settles down to a value of $\sim 0.33$ for large distances in the bulk of the system. In an infinite system, the plateau at $1/3$ will persist for $|\rho|\to\infty$; because our cylinder is finite, near the edge we observe a deviation from $1/3$. For the ground state of the Coulomb interaction, the ODLRO has small oscillations, but still appears to approach 0.33. 

Apart from the ODLRO when the system is in the Laughlin phase, we can also compute the ODLRO for the ground state perturbed away from the pure Coulomb interaction. This is conveniently done by softening the short-range component of the potential, i.e. $H_{\rm Coulomb} - \lambda H_L$. It is known that for sufficiently large $\lambda$ the ground state becomes compressible~\cite{rh85}. In Fig.~\ref{fig:odlro_n_9} we indeed see that the ODLRO drops rapidly to zero as $\lambda$ is increased, indicating the destruction of the FQHE state.

\begin{figure}
\centering
\includegraphics[width=4 in]{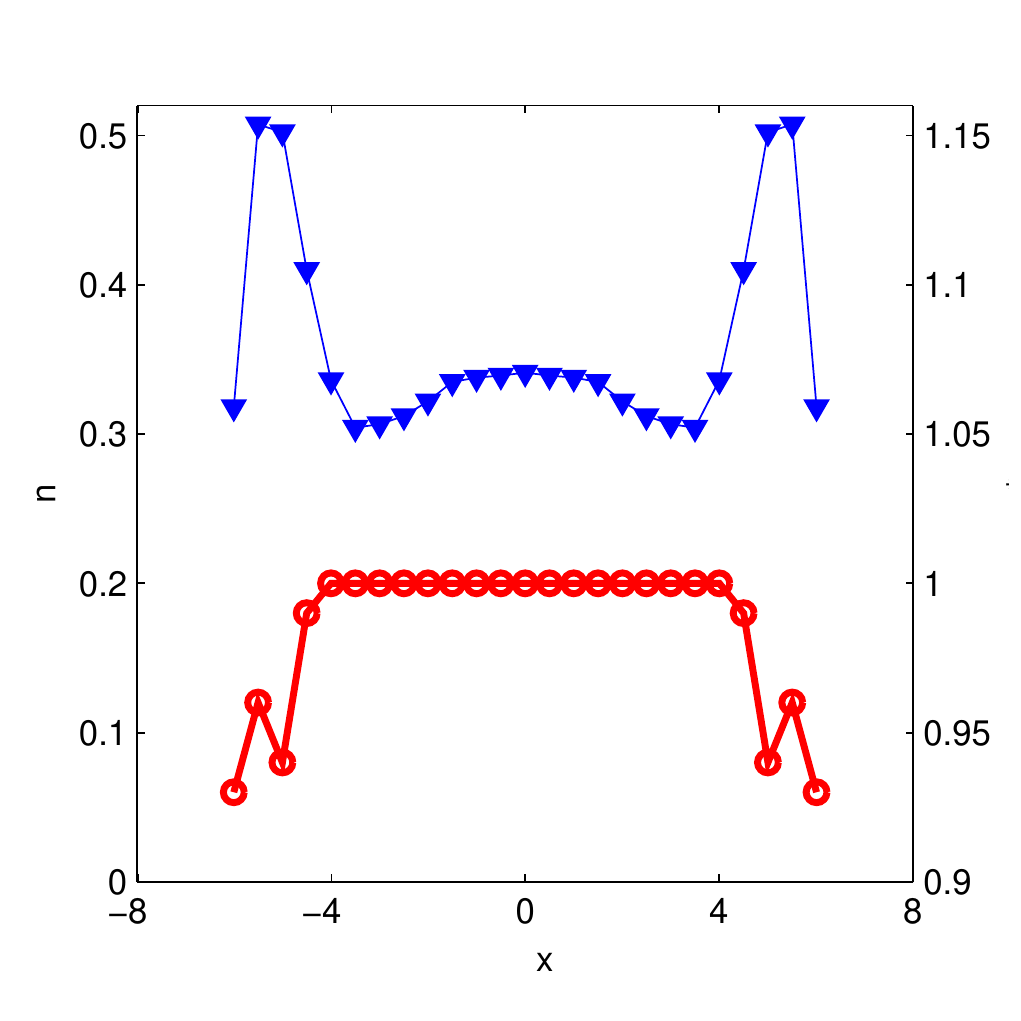}
\caption{(Color online) The occupation numbers for the Laughlin state of $N_e=9$ electrons in $N_{\rm orb}=25$ orbitals (blue, left $y$-axis), and the detected metric $g^*$ using Eq.(\ref{eq:metric_gc}) (red, right $y$-axis). Note that $g^*$ starts to deviate from the uniform value as the edge is approached from the bulk.}
\label{fig:odlro_edge_n_9}
\end{figure}

One advantage of our formulation of the ODLRO in Eq. (\ref{eq:odlro}) is that it naturally generalizes to the case of non-circular metric. Practically, the generalization of coherent states (and therefore ODLRO) for the non-Euclidean metric is done by redefining
\begin{eqnarray}
q^2 \to g_{ab} q^a q^b
\end{eqnarray}
in Eq. (\ref{eq:nm}). Note that the metric is the shape of the coherent state, which does not have to be Euclidean. For example, if the effective mass of the electrons (cyclotron metric) and the dielectric constant (guiding center metric) have different forms, the resulting metric of the droplets will be a compromise between the two that minimizes the energy \cite{haldane_geometry, boyang, haowang}. In a real sample, the metric can in fact be continuously variable -- the droplets simply assume a shape that lowers the overall energy of the liquid as much as possible. 

As the guiding centers approach an edge of the system, the droplets will get squeezed, i.e., their local metric will deviate from the one in the bulk. This distortion can be measured using the operators $\hat n_m(\rho)$. For the $\nu=1/3$ Laughlin state $\Psi_0$ with Euclidean metric $g_0$ everywhere, we observe that in the bulk
\begin{eqnarray}\label{eq:metric}
\langle \Psi_0 | \hat{n}_{0,g_0} (\mathds{1}-\hat{n}_{1,g_0}) (\mathds{1}-\hat{n}_{2,g_0})   \hat{n}_{0,g_0} |\Psi_0  \rangle \approx 1.
\end{eqnarray}
This means that once we create an electron in the coherent state $m=0$ around some point, the physics of the Laughlin state takes care of preventing the occupation of the next two orbitals $m=1,2$. In the bulk of the exact Laughlin state, the overlap in Eq. (\ref{eq:metric}) is exactly equal to one; in the Coulomb ground state, it is close but not strictly equal to one because other electrons still have a small amplitude to enter the droplet already containing one electron.

With this in mind, we propose that the optimum metric at some point $\rho$ is given by varying the metric $g$ so that the overlap of $(\mathds{1}-\hat{n}_{2,g})(\mathds{1}-\hat{n}_{1,g})\hat{n}_{0,g}\Psi_0$ and $\hat{n}_{0,g}\Psi_0$ is maximized: 
\begin{equation}\label{eq:metric_gc}
g^*(\rho) = \min_g \{ |\langle \Psi_0 | \hat{n}_{0,g_0}(\rho) (\mathds{1}-\hat{n}_{1,g_0}(\rho))(\mathds{1}-\hat{n}_{2,g_0}(\rho)) | \hat{n}_{0,g_0}(\rho)| \Psi_0 \rangle - 1| \}.
\end{equation}
In Fig. \ref{fig:odlro_edge_n_9}, we show the results for the optimum metric $g^*$ obtained in this way for the Laughlin wave function for a system of $N_e=9$ electrons. We make the approximation of keeping the off-diagonal element $g_{xy}$ of the metric $0$ and only vary $g_{xx}$ ($g_{yy}=g_{xx}^{-1}$). It is clear that the metric which determines the shape of the droplets deviates from $1$ as the edge is approached from the bulk of the fluid. With this method, we are able to obtain a spatial map of the droplets as their shape varies throughout the fluid. Fig. \ref{fig:odlro_edge_n_9} also shows that the deviation in the metric near the edge is accompanied by a deviation in the electron occupation numbers from the bulk filling factor. In the next section, we introduce the theoretical prediction which relates the occupation numbers to the second gradient of the metric [Eq. \ref{gdensity}] in the bulk of the system. 

\section{Metric Perturbation}\label{sec:response}

In the previous Section, we studied how the guiding-center metric varies over the surface of a system with open boundaries. In this section, we focus on the bulk of the system and vary the metric in a controlled way to measure the response of the $\nu=1/3$ state. 
\begin{figure}[htb]
\centering
\includegraphics[width=0.6\columnwidth]{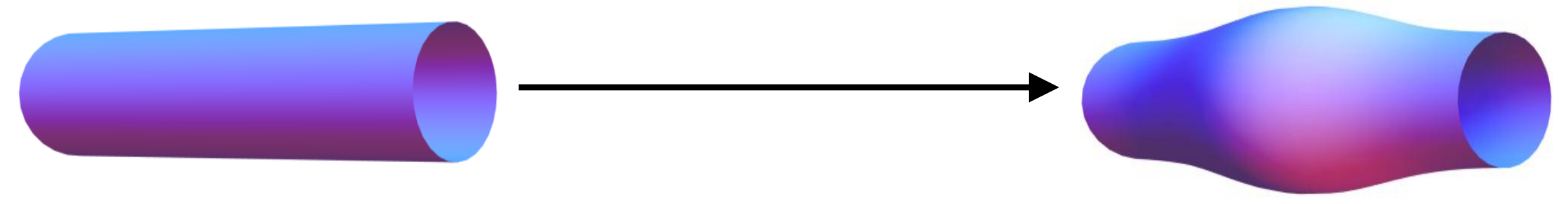}
\caption{(Color online) We perturb the system by a spatially-dependent metric deformation [Eq. (\ref{eq:metric_vary})] which can be viewed as distorting the cylinder.}
\label{fig:distort}
\end{figure}

We generalize the $V_1$ Hamiltonian to accommodate a smoothly varying diagonal metric $g={\rm diag}\left[ g_{xx}, g_{xx}^{-1}\right]$:
\begin{eqnarray}
H_L(g)=\sum_{m_1,m_2,m_3,m_4} V_{m_1,m_2,m_3,m_4}(g)c^{\dagger}_{m_1}c^{\dagger}_{m_2}c_{m_3}c_{m_4},
\end{eqnarray}
where $V_{m_1,m_2,m_3,m_4}=\langle m_1,m_2|H_L|m_3,m_4\rangle\sim \exp\bigg(-\frac{\beta^2+\gamma^2}{2g_{xx}}\bigg)\bigg(\frac{\gamma^2}{g_{xx}}-\frac{\beta^2}{g_{xx}}-1\bigg)$ ($\beta$ and $\gamma$ are defined as in Section 2). We perturb the metric as
\begin{equation}\label{eq:metric_vary}
g_{xx}=1 + \delta g, \;\;\; \delta g = a_g\exp(-b_gx^2) \ll 1,
\end{equation}
keeping the off-diagonal elements zero. Parameters $a_g,b_q$ are to be considered small, so that $g_{xx}$ varies smoothly and remains close to 1. This is equivalent to locally distorting the cylindrical surface as shown in Fig. \ref{fig:distort}. We make the simplifying assumption that $V_{m_1,m_2,m_3,m_4}(g)\sim V_{m_1,m_2,m_3,m_4}(g_M)$, where $g_M$ is the metric at the center of mass of a pair of scattering particles.

\begin{figure}[htb]
\centering
\includegraphics[width=\columnwidth]{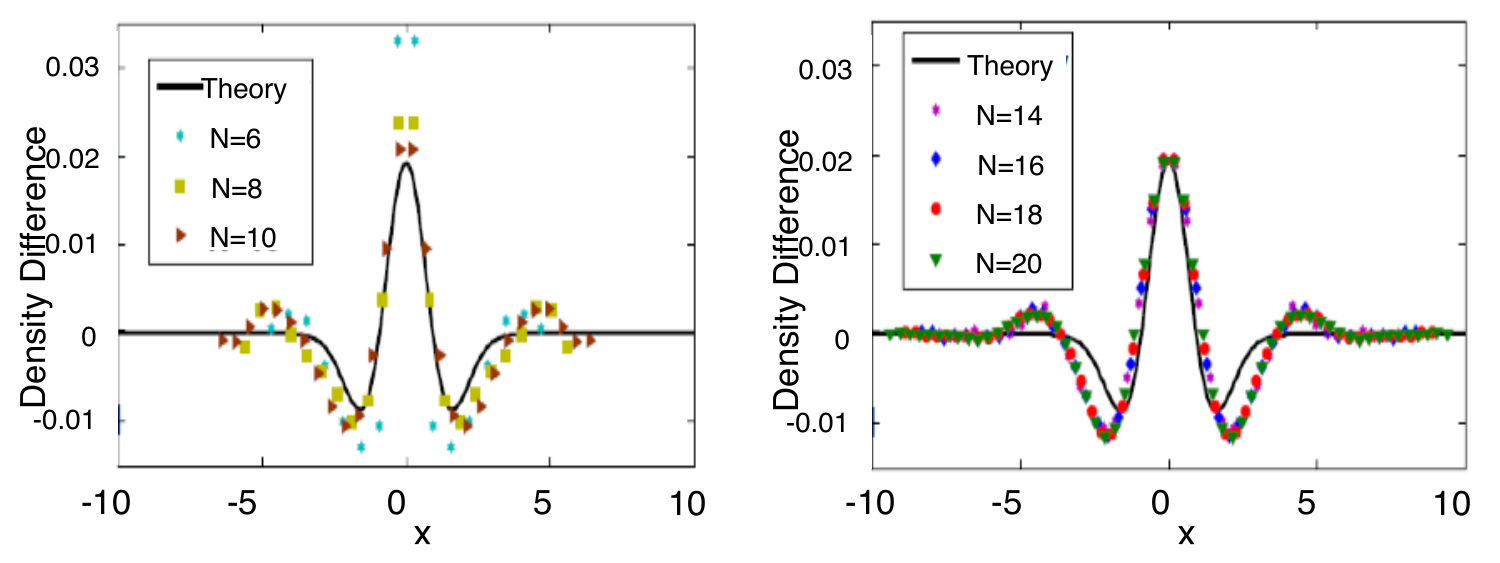}
\caption{(Color online) Response of the Laughlin state to a bulk metric perturbation. The plot shows comparison between theoretical prediction and numerical data for the difference in occupation numbers, $\delta n =n_m(a_g=0.05,b_g=0.5785)-n_m(a_g=-0.05,b_g=0.5785)$. The left panel shows the results obtained using exact diagonalization, while the right panel shows the results for larger system sizes obtained using DMRG (maximum number of kept states is 2500, discarded entropy is $10^{-12}$). The center of the cylinder is at $x=0$.}
\label{fig:density_diff}
\end{figure}
We want to compare the actual response of the fluid density, that is, the metric-dependent occupation numbers $n_m(a_g,b_g)$ to the theoretical prediction \cite{haldane_geometry} (see also Ref. \cite{bo_thesis})
\begin{equation}\label{gdensity}
\delta\rho(r)=\frac{s}{2}\partial_a\partial_b\delta g(r)^{ab},
\end{equation} 
where $s=-1$ is the guiding center spin for the Laughlin state. We use DMRG to obtain the occupation numbers for systems with the spatially varying metric as in Eq. (\ref{eq:metric_vary}). We subtract the densities obtained for positive and negative values of $a_g$ in order to double the response [Eq. (\ref{gdensity})] and thus make it easier to measure. In Fig. \ref{fig:density_diff}, we show the results for the density difference between a system with $a_g=0.05$ and $a_g=-0.05$ and compare with the theoretical result in Eq. (\ref{gdensity}). We see that the numerics have converged everywhere by reaching the system size $N_e=20$ electrons. 
We note that theory accurately reproduces the numerical data to the leading order, although some visible deviations exist in the form of secondary peaks away from the perturbation site.  Those features are not captured by Eq. (\ref{gdensity}) but we expect them to accounted for by $O(\delta g^2)$ terms. Given the near convergence of the numerical results for larger sizes, a more complete calculation of the density fluctuation extending beyond linear response merits further investigation. 

\begin{figure}[htb]
\centering
\includegraphics[width=4 in]{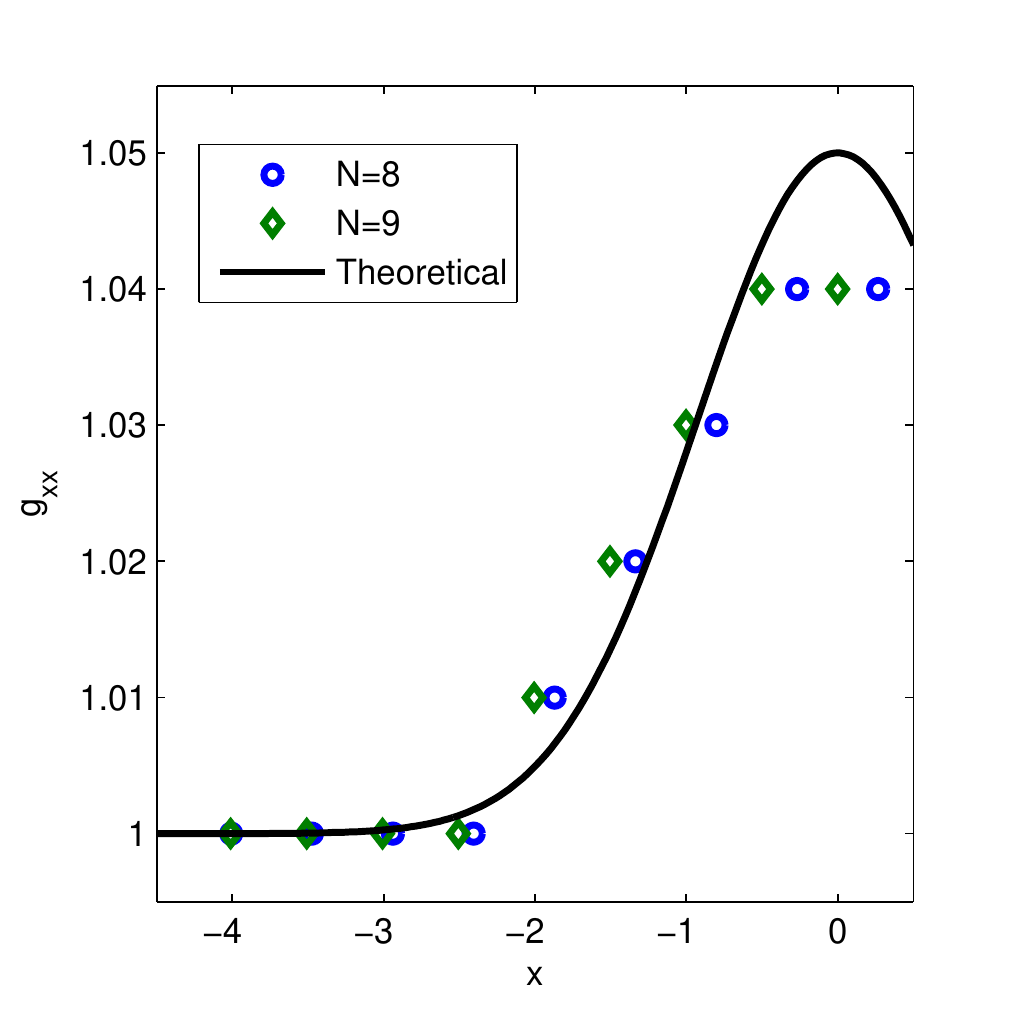}
\caption{(Color online) Detected local metric (symbols) using the method in Eq. (\ref{eq:metric_gc}) compared to the actual background metric (black line) for a system with $N=8$, $9$ electrons on the cylinder. The center of the cylinder is at $x=0$.}
\label{fig:met_det}
\end{figure}

We can also use the metric-detection technique we developed in the last section (Eq. \ref{eq:metric_gc}) to check if we can deduce the spatial variation of the metric from the wave function. Fig. \ref{fig:met_det} shows the optimum metric obtained by minimizing the value of the right hand side of Eq. \ref{eq:metric_gc}  for a system of $N=8$ and $9$ electrons. The metric detected by this method from the ground state of the Hamiltonian in Eq. (\ref{eq:metric_gc}) corresponds closely to the input metric in Eq. \ref{eq:metric_vary}. The deviation near the center ($x=0$) is attributable to the limited resolution of a finite system. 

\section{The pair amplitude operator}\label{sec:pair}

In this Section we demonstrate a different method to measure the metric of the Laughlin state based on the pair creation operators first introduced in Ref.~\cite{haldane_book}. The method is formulated in orbital space and therefore applies predominantly to the bulk of the fluid.

The pair creation operator is an operator that creates two particles in a state of relative angular momentum $M$. On a cylinder in the Landau gauge, angular momentum is not a good quantum number, but $M$ can be interpreted as the average separation between particles forming a pair.  
The simplest examples are momentum $M=0$:
\begin{equation}\label{eq:p0}
\hat{P}^{\dagger}_{M=0}(p) = \sum_r e^{-X^2_r} c^{\dagger}_{p+r}c^{\dagger}_{p-r}.
\end{equation}
and momentum $M=1$:
\begin{equation}\label{eq:p1}
\hat{P}^{\dagger}_{M=1}(p) = \sum_r X_r e^{-X^2_r} c^{\dagger}_{p+r}c^{\dagger}_{p-r}.
\end{equation}
Here $p$ is fixed and labels the center of mass of a pair of particles that are being created ($c^{\dagger}$ are the usual electron creation operators). We have also introduced $X_r \equiv 2\pi r/L$. Note that $p\pm r$ must be integers as they label the single particle orbitals, which means that $p$ and $r$ can assume integer or half-integer values. 

The physical significance of operators (\ref{eq:p0}) and (\ref{eq:p1}) is that they form ``one half" of the parent Hamiltonians of the $\nu=1/3$ Laughlin state:
\begin{equation}
H_{Laughlin, Bose}=\sum_p \hat{P}^{\dagger}_0 (p) \hat{P}_0 (p), \;\;\; H_{Laughlin, Fermi}=\sum_p \hat{P}^{\dagger}_1 (p) \hat{P}_1 (p).
\end{equation}
The role of $\hat{P}^{\dagger}_1$ is to create a pair of particles with relative angular momentum 1. Each such pair is assigned a positive energy penalty due to the term $P_1^\dagger P_1$ at all possible values of $p$. Therefore, any pair with relative angular momentum$=1$ is assigned an energy of the order $1$, and the resulting Laughlin state (which is a zero-energy ground state of the Hamiltonian) ends up having no pairs of particles in a state with relative angular momentum $1$. In the rest of this Section, we focus on the fermionic states for which only odd values of $M$ are meaningful because of Fermi statistics. 

Now imagine the opposite situation when we start from an unknown state $\Psi$, and compute
\begin{equation}
\langle\Psi|\hat{P}^{\dagger}_1\hat{P}_1|\Psi\rangle
\end{equation}
for all $p$. If we find this amplitude to be zero for any $p$, the unknown state at $1/3$ must be the Laughlin state as it is the only state that has exactly zero amplitude for all pairs in the relative momentum $M=1$ state.
Therefore, we can refer to this ``pair amplitude being zero in the momentum channel $M=1$" as the definition of the Laughlin state.

It is straightforward to generalize the above to any momentum channel $M$:
\begin{equation}
\hat{P}^{\dagger}_M(p)=\sum_r \frac{1}{\sqrt{2^M M!}}H_M(X_r\sqrt{2})e^{-X_r^2}c_{p+r}^\dagger c_{p-r}^\dagger,
\end{equation}
where $H_M$ is the Hermite polynomial. The normalization of the operators $\hat{P}_M(p)$ is fixed by demanding that the eigenvalues of $\hat{P}^{\dagger}_M(p)\hat{P}_M(p)$ for $2$ particles in a large number of orbitals are only 0 or 1, for any value of $p$ in the bulk of the system. We also perform some consistency checks for the numerical implementation of operators $\hat{P}_M$. For example, if we compute $\eta_M = \langle\Psi|\hat{P}^{\dagger}_M(p)\hat{P}_M(p)|\Psi\rangle$ for the Laughlin state $\Psi= \Psi_L$, we must
find 
\begin{equation}
\eta_0 = 0, \eta_1 = 0, \eta_2 = 0, \eta_3 > 0, \eta_4 = 0, \eta_5 > 0 \ldots
\end{equation}
i.e., all even pair amplitudes must vanish because of Fermi antisymmetry, and among the odd ones $\eta_1$ also vanishes (which is the special property of the Laughlin state), but higher odd ones ($\eta_5$, $\eta_7$, etc.) are in general non-zero. If we take instead $\Psi$ to be the Coulomb ground state at $\nu=1/3$ (with a large overlap with the Laughlin state), $\eta_1$ is no longer strictly zero, but it is still much smaller $\eta_1 \ll \eta_3, \eta_5, \ldots$~\cite{haldane_book}.

Having introduced the operators $\hat{P}_M$, we now show that they can be used to measure the intrinsic geometry of any state. The pair amplitude operators are more convenient than ODLRO operators because they respect the symmetry of the system, i.e., we can restrict to blocks of the Hilbert space corresponding to fixed total momentum along the cylinder. 
Similarly to ODLRO operators, the next step is to generalize $\hat{P}_M$ to measure the intrinsic geometry of a state at various points $p$. The generalized pair amplitude operator for a general diagonal metric $g={\rm diag}\left[\alpha,\alpha^{-1}\right]$ is given by 
\begin{eqnarray}
\hat{P}_{M,\alpha} (p) = \sum_{r}\frac{1}{\sqrt{\sqrt{\alpha}2^M M!}} H_M(X_r\sqrt{2/\alpha}) e^{-X_r^2/\alpha} c_{p+r}c_{p-r} 
\end{eqnarray}
Notice that the dominant effect of the metric is to ``squeeze" the Gaussian factor, which is similar to varying the aspect ratio of the cylinder. One can easily verify that this expression reduces to the correct one for $M=1$, describing the Hamiltonian for the anisotropic Laughlin state~\cite{boyang}: 
\begin{eqnarray}\label{eq:laughlin_anis}
\nonumber V_{m_1m_2m_3m_4} &=& \frac{1}{rN_{\rm orb}\alpha^3}e^{-\frac{1}{2\alpha} \left[ \left(X_{m_1}-X_{m_3}\right)^2 +\left(X_{m_1}-X_{m_4}\right)^2\right]} \\
&& \times \left[ \left(X_{m_1}-X_{m_3}\right)^2 - \left(X_{m_1}-X_{m_4}\right)^2 \right].
\end{eqnarray}

\begin{figure}
\centering
\includegraphics[width=4 in]{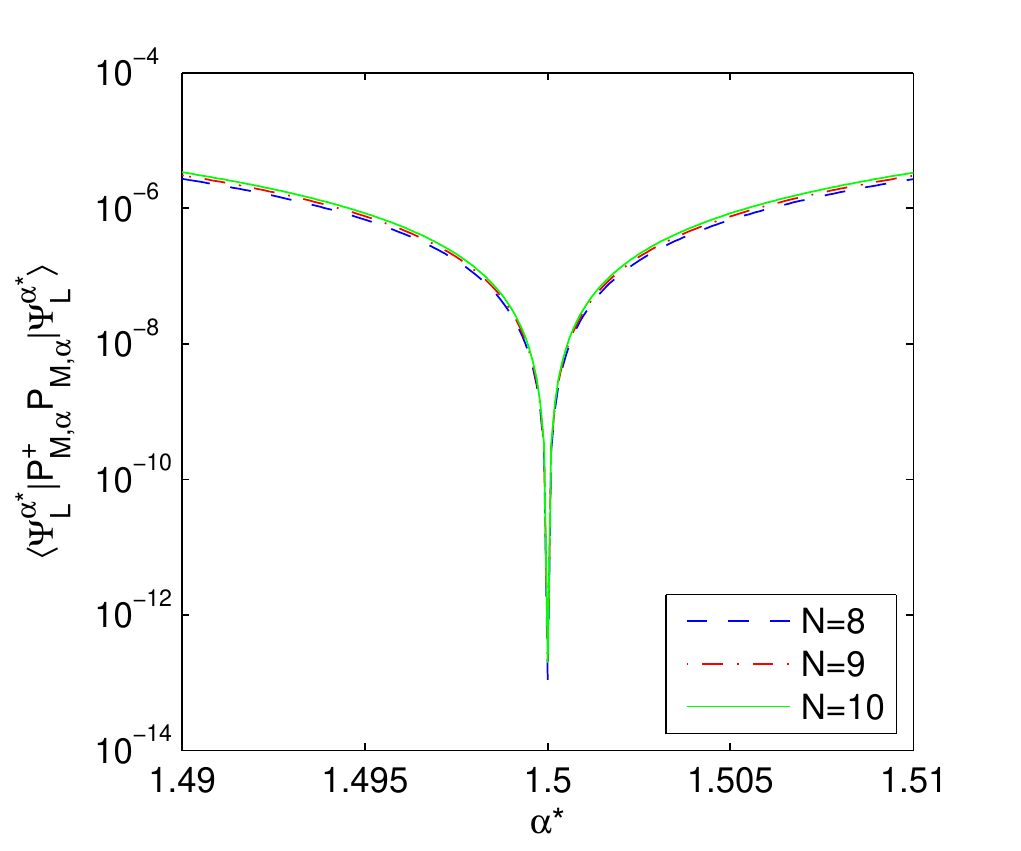}
\caption{(Color online) The expectation value of the pair amplitude operator detects a correct value of the anisotropy by the sharp minimum at $\alpha=\alpha^*=1.5$ and $M=1$.}
\label{fig:pairamp_delta_anis}
\end{figure}

We are now in position to demonstrate how the operator $\hat{P}_{M,\alpha}$ can be used to measure the intrinsic anisotropy of a state. As a consistency check, we first measure the geometry of an anisotropic Laughlin state obtained by diagonalizing the Hamiltonian in Eq. (\ref{eq:laughlin_anis}). We fix a value $\alpha^*=1.5$ and indeed find that the Hamiltonian (\ref{eq:laughlin_anis}) has a zero-energy ground state $\Psi^{\alpha^*}_L$. Next, we take this $\Psi^{\alpha^*}_L$, and evaluate the expectation value 
\begin{eqnarray}
\langle\Psi^{\alpha^*}_L|\hat{P}^{\dagger}_{1,\alpha}(p)\hat{P}_{1,\alpha}(p)|\Psi^{\alpha^*}_L\rangle.
\end{eqnarray}
That is, we compute the expectation value of the pair amplitude $M=1$ operator in the previously obtained ground state. This expectation value is a function of two parameters, $p$ and $\alpha$. For simplicity, we fix $p$ to be in center of the cylinder (corresponding to the bulk of the system), and vary $\alpha$. The expectation value as a function of $\alpha$ is plotted in Fig. \ref{fig:pairamp_delta_anis}. The plot shows a deep minimum for exactly $\alpha = \alpha^*$. This calculation demonstrates that our pair amplitude operator has detected that the intrinsic metric of the state is exactly the one given by the anisotropy that was explicitly used as an input for the calculation.

\begin{figure}
\centering
\includegraphics[width=4 in]{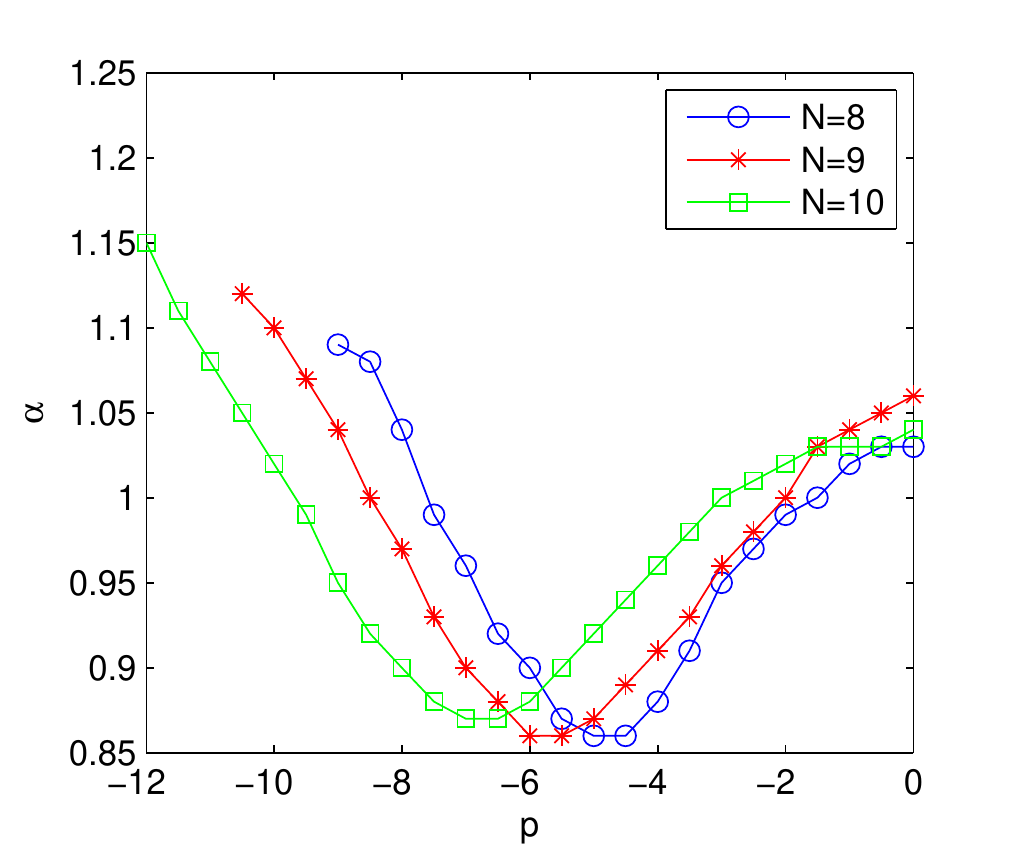}
\caption{(Color online) Change in the detected metric $\alpha$ going from the edge to the center for a system of $N=8$, $9$, and $10$ electrons in $N_{\rm orb}=3N-2$ orbitals at $\nu=1/3$ with Coulomb interaction.}
\label{fig:pairamp_coulomb}
\end{figure}

After this necessary consistency check, we can study more complicated cases, for example the Coulomb interaction at $\nu=1/3$. We compute  $\langle\Psi^{\alpha^*}_L|\hat{P}^{\dagger}_{1,\alpha}\hat{P}_{1,\alpha}(p)|\Psi^{\alpha^*}_L\rangle$as we change $p, \alpha$. The results are shown in Fig. \ref{fig:pairamp_coulomb}. We see that even though the Coulomb interaction is isotropic, because we have an open cylinder, there are fluctuations in the metric of the state as we move $p$ along the axis of the cylinder. In this case even the bulk metric is not fully isotropic because of the finite size of the system and the long-range nature of the interaction potential. As we mentioned earlier, since this method is defined in orbital space, it is not expected to work correctly at the edge of the system because of strong constraints on the occupation of one-body orbitals arising from the Jack polynomial structure of the Laughlin state~\cite{jack}. This issue does not arise for fully periodic boundary conditions (torus). 

\begin{figure}
\centering
\includegraphics[width=4 in]{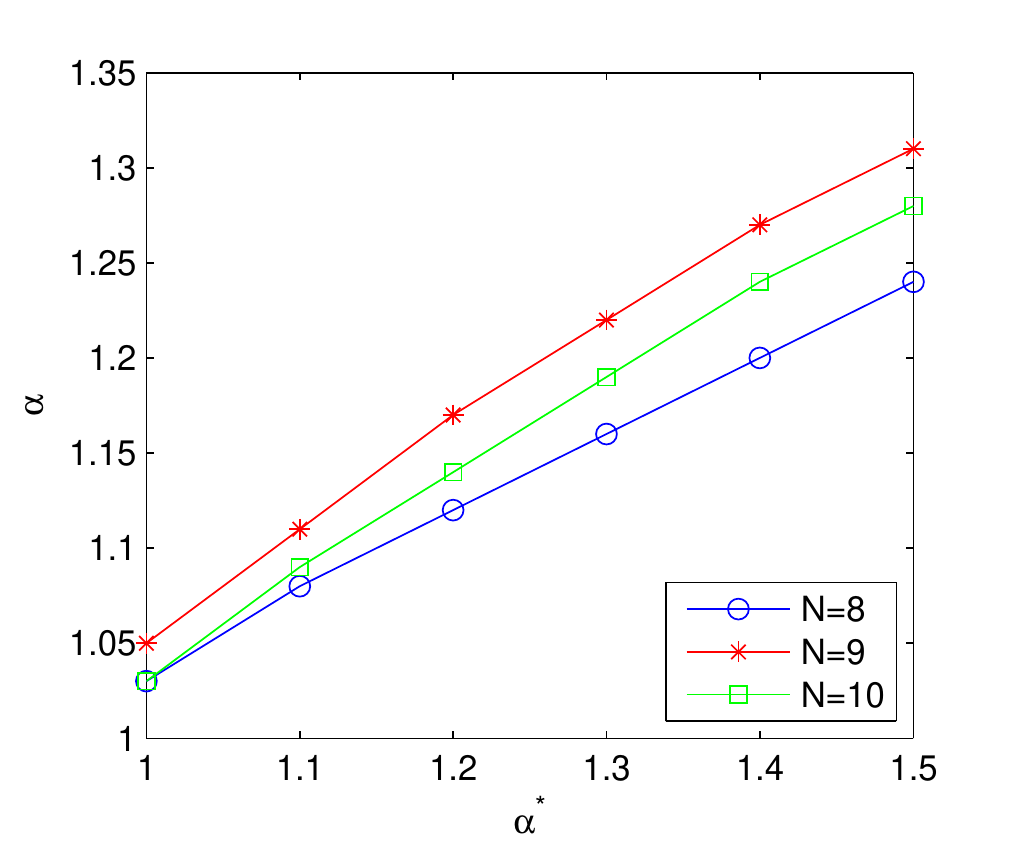}
\caption{Detected metric $\alpha$ as a function of the mass metric $\alpha^*$ for a system of $N=8$, $9$, and $10$ electrons at $\nu=1/3$ with Coulomb interaction.}
\label{fig:pairamp_coulomb_anis}
\end{figure}

Next, we consider Coulomb interaction combined with mass anisotropy. By this we mean that the mass anisotropy appears in the form factor resulting from single-particle wave functions, but the interaction Fourier transform, $V(q)$, remains isotropic [similar to Refs.~\cite{boyang, haowang}]. Now that there is a mismatch between the metric in $V(q)$ and the Gaussian envelope $e^{-q'^2/2}$, the state needs to optimize between them. Similar to our previous experiment, we fix $\alpha^*$ in the above, and use our operator to see what metric the state itself will pick. Fig. \ref{fig:pairamp_coulomb_anis} shows  a plot of the detected metric in the bulk versus the input metric. In agreement with earlier results \cite{boyang}, we find the optimal metric lies between the mass metric and the interaction metric, though the exact values slightly differ because of different boundary conditions.

\section{Conclusions}\label{sec:conc}

We have used various numerical techniques to characterize the intrinsic metric of the $\nu=1/3$ Laughlin state based on its description in terms of particle-hole composites [Fig. \ref{fig:laughlindroplets}]. We have shown that this picture allows one to define an ODLRO for the Laughlin state, which is also sensitive to local deviations of the metric near the edge of the system. Furthermore, using the pseudopotential Hamiltonian of the Laughlin state, we have measured the response of the Laughlin state to the smooth deformations of the background metric in which it is embedded, finding good agreement with analytical expectations. Finally, we have introduced a generalized pair-amplitude operator and showed that it can be used to detect the metric of the Laughlin state when the host system has anisotropic band mass. Our calculations have been implemented in the cylinder geometry, but they are sufficiently general and directly apply to incompressible states at other filling fractions and with other types of boundary conditions (for example, disk or torus). 

As we mentioned in the Introduction, the fundamental role of quantum geometry in FQHE has been the subject of several recent theoretical papers \cite{haldane_geometry, wiegmann, gromov1, bradlyn}. Moreover, recent experiments on GaAs quantum wells \cite{shayegan, shayegan2, shayegan3} have measured the anisotropy of the Fermi contour in the case of the compressible $\nu=1/2$ state. In these experiments, anisotropy is induced by the tilting of the magnetic field, which is unfortunately rather complicated to model theoretically~\cite{papic_tilted}. A simpler way to induce anisotropy is to vary the band mass tensor~\cite{boyang, haowang}. This may be relevant for certain materials like AlAs, but inducing \emph{local} metric variations in this way may still prove challenging. We note that non-uniform in-plane electric field has the effect of changing the metric in qualitatively the same way as mass anisotropy [see Appendix B]. Therefore, applying the gradient of an electric field may be a more convenient way in practice to measure the local density response of FQH states to a non-uniform metric.   

\section{Acknowledgments}

This work was supported by DOE grant DE-SC0002140. FDMH also acknowledges support from the W. M. Keck Foundation.

\appendix 

\section{Effect of cylinder circumference}

The circumference $L$ of the cylinder is a non-trivial tunable parameter describing the system. It sets the distance, $2\pi/L$, between the centers of orbitals along the $x$-axis, which has an effect on the interaction matrix elements. Additionally, this will also determine the number of orbitals which are affected by the edge. 

\begin{figure}[htb]
\centering
\includegraphics[width=0.5\columnwidth]{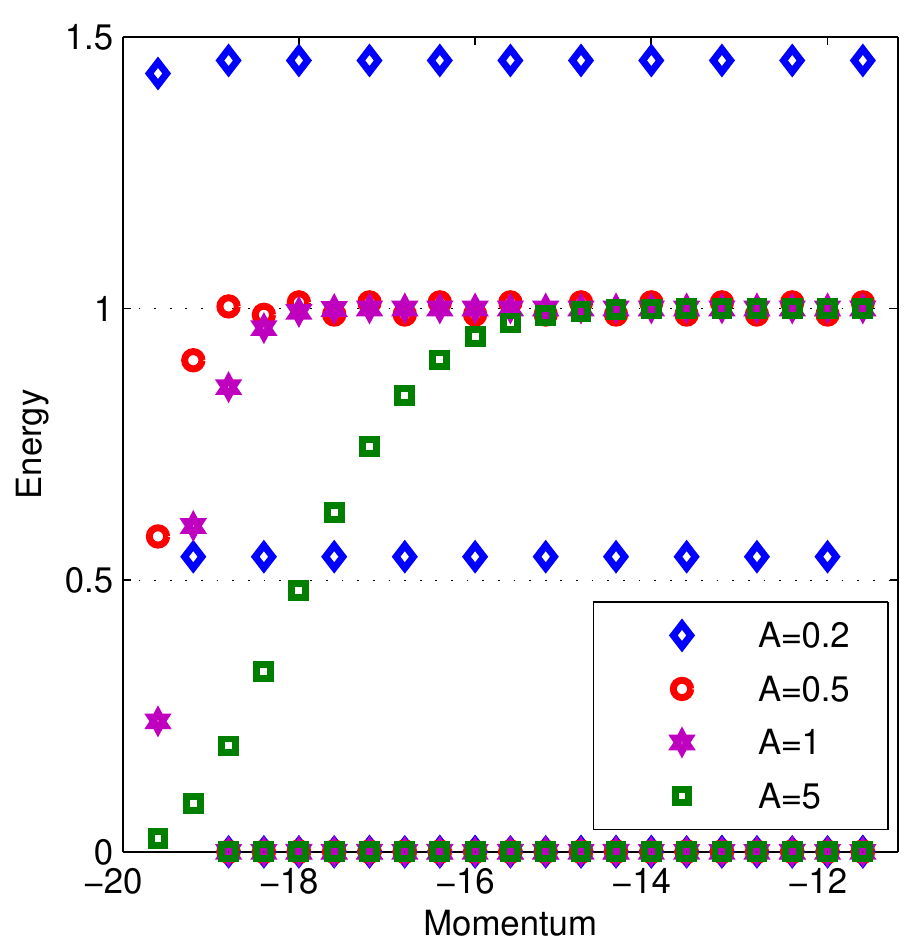}
\caption{Spectrum for $H_L$ using $N=2$ electrons in $N_{\rm orb}=20$ orbitals as a function of their combined momentum for different values of the aspect ratio, $A=L/H$. The orbitals are labelled from $m=-10$ to $m=10$. Note that the zero energies have large degeneracies but the non-zero energies are non-degenerate.}
\label{fig:N_2}
\end{figure}

To see this, let us start with a toy system consisting of only two electrons on the cylinder. The spectrum for the short-range $H_L$ is shown in Fig. \ref{fig:N_2} for different values of the aspect ratio, $A=L/H$. At each value of the momentum in an infinite cylinder, we would expect one eigenvalue at non-zero energy (corresponding to the electrons in nearest or next-nearest orbitals) and the rest to be zero. 

For $A=1$, the non-zero eigenvalue is constant in the bulk but approaches zero at momenta which place both electrons near the edge of the system. The number of orbitals for which the energy deviates from its constant value in the bulk can be taken as an indicator of the penetration depth of the edge effect. For $A>1$, more orbitals are affected by the edge because the distance between orbitals ($2\pi/L$) decreases. For $A<1$, the non-zero eigenvalue starts to oscillate between two fixed values. [ See, in particular, the data for A = 0.2 in Fig. A1]. This indicates that the symmetry of a ground state with several electrons will also change from that of a uniform liquid for $A\ll 1$.

\begin{figure}[htb]
\centering
\includegraphics[width=0.5\columnwidth]{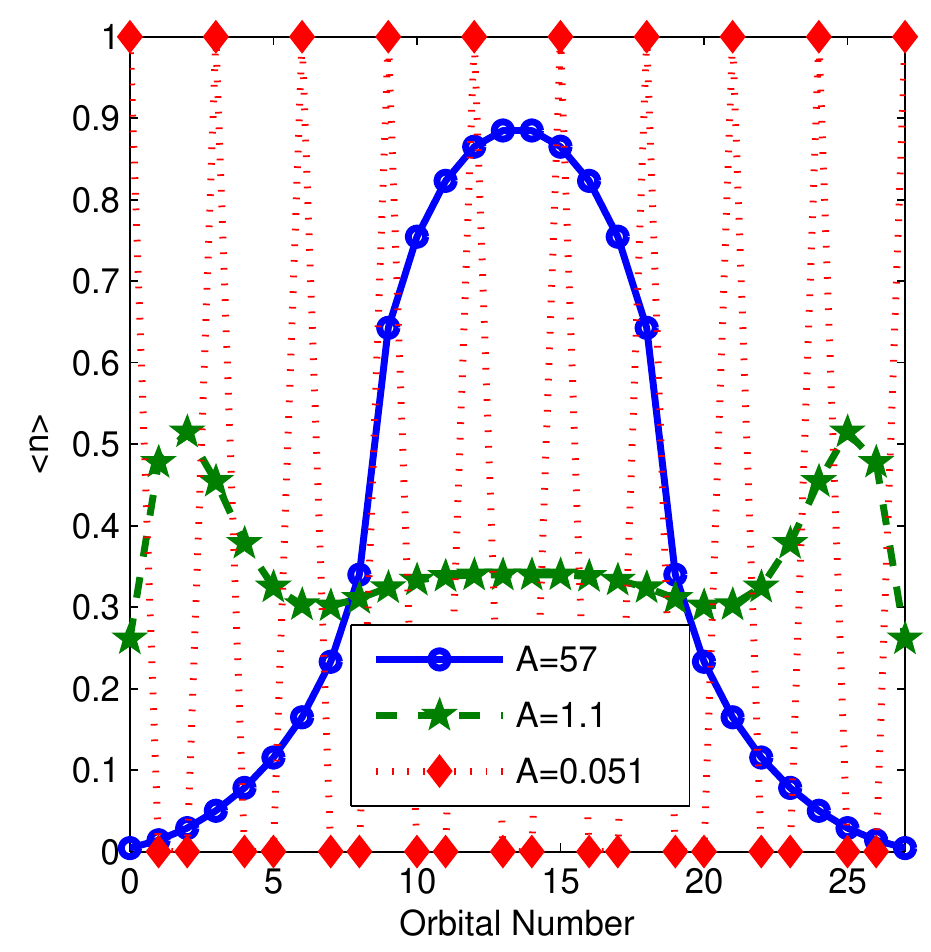}
\caption{Occupation numbers for $N=10$ electrons in $N_{\rm orb}=28$ orbitals on the cylinder for different values of the aspect ratio, $A=L/H$, ($A\gg 1$, $A\approx 1$ and $A \ll1$).}
\label{fig:phases_L}
\end{figure}

Previous work has shown that in the extreme limit of $A\ll 1$ and $A \gg 1$, the FQH liquid is no longer possible~\cite{rezayi_cylinder}. This is intuitively clear since the FQHE is a two-dimensional phenomenon. We show the different natures of the ground state at extreme values of the order parameter in Fig. \ref{fig:phases_L}. When $L$ is very small, we are in the Tao-Thouless limit\cite{tao-th1, tao-th2} and obtain a one-dimensional charge density wave. The leading order of the electron-electron interaction is then electrostatic in nature.

When $L$ is large, we reach the conformal limit, in which the wave function can be defined through the coefficients of the Jack polynomials \cite{jack} alone without requiring any surface-dependent normalization factors. This limit is simple for analytical calculations; however, it does not give the right occupation numbers in the bulk. Only when $A$ is close to 1 can a quantum Hall liquid be recognized with occupation numbers close to $1/3$ in the bulk.

Since the Hall effect is essentially a two dimensional phenomenon, confining to one dimension as in the hoop limit or the Tao-Thouless limit does not give the right physics. The ground state in the Tao-Thouless limit breaks translational symmetry. While the hoop limit ground state does not break translational symmetry, all geometrical information is lost and there is no meaning to individual orbitals any more since they are all on top of each other. Therefore several of its properties are also different from a usual FQHE state \cite{soule}. 

\section{Electric field gradient}

Here we show that the gradient of an electric field can be used to induce the anisotropic metric in a way that is very similar to the the mass anisotropy. 

The one-body Hamiltonian with the electric field gradient along $x$-axis and strength $\alpha$ is
\begin{equation}
H=\frac{p_x^2}{2m}+\frac{(p_y+eBx)^2}{2m}+\frac{1}{2}\alpha x^2.
\end{equation}
Without any electric field, it is well-known that the Hamiltonian maps to the harmonic oscillator problem
\begin{eqnarray}
H=-\frac{\hbar^2}{2m}\frac{d^2}{dx^2}+\frac{1}{2}m\omega_c^2 \left( x + k_y \ell_B^2\right)^2.
\end{eqnarray}
When $\alpha\neq 0$, we can still express the Hamiltonian in the above form by rescaling the cyclotron energy and the magnetic length:
\begin{eqnarray}
\tilde\omega_c^2 =\omega_c^2 (1+a), \;\; \tilde \ell_B^2 = \frac{\ell_B^2}{1+a}, \;\; a=\frac{\alpha}{m\omega_c^2}. 
\end{eqnarray}
The new eigenvalues acquire a dispersion
\begin{eqnarray}
E_{n,k_y}=\hbar \omega_c \sqrt{1+a} (n+1/2) + \frac{\hbar \omega_c (k_y \ell_B)^2 a}{2(1+a)},
\end{eqnarray}
and the eigenvectors are 
\begin{eqnarray}
\phi_{n,k_y}=\sqrt{\frac{1}{L \frac{\ell_B\sqrt{\pi}}{\sqrt{1+a}} 2^n n! }} e^{i k_y y-\frac{1+a}{2\ell_B^2} \left( x + \frac{k_y \ell_B^2}{(1+a)} \right)^2   }H_n \left( \sqrt{1+a} ( x/\ell_B + \frac{k_y\ell_B}{1+a} ) \right),
\end{eqnarray}
where $k_y=2\pi j/L$, $j=-N_{\phi}/2,\ldots,N_{\phi}/2$.

The interaction matrix element for the Laughlin $V_1$ interaction is given by
\begin{eqnarray}
\nonumber V_{j_1j_2j_3j_4} &=& \frac{\sqrt{1+a}}{\sqrt{rN_{orb}}} e^{-\frac{1}{2(1+a)}\left( (X_{j1}-X_{j_3})^2 + (X_{j_1}-X_{j_4})^2 \right)} \\
&& \left[ (X_{j_1}-X_{j_3})^2-(X_{j_1}-X_{j_4})^2 - (1+a) \right],
\end{eqnarray}
where $X_j=2\pi j/L$,$N_{orb}=N_{\phi}+1$, and $r$ denotes the aspect ratio ($L^2=2\pi r N_{orb}$).

We see that the leading order effect of the electric field gradient is coming from the denominator of the 
Gaussian and is very similar to mass anisotropy [compare with Eq. (\ref{eq:laughlin_anis})]. The two effects are of course not identical because of the extra single-particle terms in the electric field case. However, one can empirically establish that the effect of extra terms is small. We have verified that the method of pair amplitude operators detects roughly the same value of the input electric field, similar to the anisotropy calculation in Fig. \ref{fig:pairamp_delta_anis}.


\section*{References}

\begin{thebibliography}{99}

\bibitem{tsui_prl} D. C. Tsui, H. L. Stormer, and A. C. Gossard, Phys. Rev. Lett. {\bf 48}, 1559 (1982).

\bibitem{stormer_rmp}  H. L. Stormer, Rev. Mod. Phys. {\bf 71}, 875 (1999).

\bibitem{laughlin} R. B. Laughlin, Phys. Rev. Lett. {\bf 50}, 1395 (1983).

\bibitem{wen_book} X. G. Wen, \emph{Quantum Field Theory of Many-Body Systems}, Oxford University Press, 2004.

\bibitem{arovas_fractional}  D. Arovas, J.R. Schrieffer and F. Wilczek, Phys. Rev. Lett. {\bf 53}, 722 (1984).

\bibitem{prange}
\emph{The Quantum Hall Effect}, 2nd ed., edited by R. E. Prange and S. M. Girvin, Springer-Verlag, New York, 1990.


\bibitem{qh_book} T. Chakraborty and P. Pietilainen, \emph{The Quantum Hall Effects, Fractional and Integral}, Springer-Verlag, Berlin, New York, 1995.

\bibitem{girvin_review} Steven Girvin, \emph{The Quantum Hall Effect: Novel Excitations and Broken Symmetries}, in A. Comtet, T. Jolicoeur, S. Ouvry, and F. David, editors, \emph{Topological Aspects of Low Dimensional Systems}, Addison Wesley, 2000.


\bibitem{jainbook} J. K. Jain, \emph{Composite fermions}, (Cambridge University Press, 2007).


\bibitem{haldane_geometry} F. D. M. Haldane, Phys. Rev. Lett. {\bf 107}, 116801 (2011).

\bibitem{zhk} S. C. Zhang, T. H. Hansson, and S. Kivelson, Phys. Rev. Lett. {\bf 62}, 82 (1989). 

\bibitem{jain89} J. K. Jain, Phys. Rev. Lett. {\bf 63}, 199 (1989).


\bibitem{wen_zee} X. G. Wen and A. Zee, Phys. Rev. Lett. {\bf 69}, 953 (1992).


\bibitem{klitzing_prl}  K. von Klitzing, G. Dorda, and M. Pepper, Phys. Rev. Lett. {\bf 45}, 494 (1980).
\bibitem{klitzing_rmp} K. von Klitzing, Rev. Mod. Phys. {\bf 58}, 519 (1986).

\bibitem{bo_thesis} Bo Yang, \emph{Geometric Aspects and Neutral Excitations in the Fractional Quantum Hall Effect}, PhD Thesis, Princeton University, (2013).

\bibitem{avron} J. E. Avron, R. Seiler, and P. G. Zograf, Phys. Rev. Lett. {\bf 75}, 697 (1995).

\bibitem{tokatly}  I. V. Tokatly, G. Vignale, J. Phys. C {\bf 21}, 275603 (2009).

\bibitem{read_viscosity} N. Read, Phys. Rev. B {\bf 79}, 045308 (2009).

\bibitem{haldane_viscosity} F. D. M. Haldane, arXiv:0906.1854 (2009).

\bibitem{readrezayi_viscosity} N. Read and E. H. Rezayi, Phys. Rev. B {\bf 84}, 085316 (2011).

\bibitem{yeje} YeJe Park, F. D. M. Haldane, Phys. Rev. B {\bf 90}, 045123 (2014).

\bibitem{son} Dam Thanh Son, arXiv:1306.0638 (2013).

\bibitem{gromov1} A. Gromov and A. G. Abanov, Phys. Rev. Lett. {\bf 114}, 016802 (2015).

\bibitem{bradlyn} Barry Bradlyn, N. Read, Phys. Rev. B {\bf 91}, 165306 (2015).


\bibitem{gromov2} A. Gromov and A. G. Abanov, Phys. Rev. Lett. {\bf 113}, 266802 (2014).

\bibitem{wiegmann} T. Can, M. Laskin, and P. Wiegmann, Phys. Rev. Lett.
{\bf 113}, 046803 (2014).

\bibitem{wiegmann2} T. Can, M. Laskin, P. Wiegmann, arXiv:1411.3105 (2014).


\bibitem{bradlyn2} Barry Bradlyn, N. Read, Phys. Rev. B {\bf 91}, 125303 (2015).

\bibitem{boyang}  Bo Yang, Z. Papic, E. H. Rezayi, R. N. Bhatt, and F. D.
M. Haldane, Phys. Rev. B {\bf 85}, 165318 (2012).

\bibitem{haowang} Hao Wang, Rajesh Narayanan, Xin Wan, and Fuchun Zhang, Phys. Rev. B {\bf 86}, 035122 (2012)

\bibitem{apalkov} Vadim M. Apalkov and Tapash Chakraborty, Solid State Communications {\bf 177} 128 (2014).

\bibitem{areg} Areg Ghazaryan and Tapash Chakraborty, Phys. Rev. B {\bf 92}, 165409 (2015).

\bibitem{papic_tilted} Z. Papic, Phys. Rev. B {\bf 87}, 245315 (2013).

\bibitem{maciejko} J. Maciejko, B. Hsu, S. A. Kivelson, YeJe Park, and S. L. Sondhi, Phys. Rev. B {\bf 88}, 125137 (2013).

\bibitem{kunyang1} K. Yang, Phys. Rev. B {\bf 88}, 241105 (2013).

\bibitem{kunyang2} Kun Yang, arXiv:1508.01424 (2015).

\bibitem{shayegan} M. A. Mueed, D. Kamburov, M. Shayegan, L. N. Pfeiffer, K. W. West, K. W. Baldwin, and R. Winkler, Phys. Rev. Lett. {\bf 114}, 236404 (2015).
\bibitem{shayegan2} M. A. Mueed, D. Kamburov, Yang Liu, M. Shayegan, L. N. Pfeiffer, K. W. West, K. W. Baldwin, and R. Winkler, Phys. Rev. Lett. {\bf 114}, 176805 (2015).
\bibitem{shayegan3} D. Kamburov, Yang Liu, M. Shayegan, L. N. Pfeiffer, K. W. West, and K. W. Baldwin, Phys. Rev. Lett. {\bf 110}, 206801 (2013).

\bibitem{dmrg_white} Steven R. White, Phys. Rev. Lett. {\bf 69}, 2863 (1992).

\bibitem{mps1} M. Fannes, B. Nachtergaele, and R. Werner, Communications
in Mathematical Physics {\bf 144}, 443 (1992).
\bibitem{mps2} S. Ostlund and S. Rommer, Phys. Rev. Lett. 75, 3537 (1995).
\bibitem{mps3} S. Rommer and S. Ostlund, Phys. Rev. B {\bf 55}, 2164 (1997).


\bibitem{dmrg_shibata} Naokazu Shibata and Daijiro Yoshioka, Phys. Rev. Lett. {\bf 86}, 5755 (2001).

\bibitem{dmrg_feiguin} A. E. Feiguin, E. Rezayi, C. Nayak, and S. Das Sarma, Phys. Rev.
Lett. {\bf 100}, 166803 (2008).

\bibitem{dmrg_kovrizhin} D. L. Kovrizhin, Phys. Rev. B {\bf 81}, 125130 (2010).

\bibitem{dmrg_zhao} Jize Zhao, D. N. Sheng, and F. D. M. Haldane, Phys. Rev. B {\bf 83}, 195135 (2011).

\bibitem{dmrg_cyl} Zi-Xiang Hu, Z. Papic, S. Johri, R. N. Bhatt, Peter Schmitteckert, Phys. Lett. A {\bf 376}, 2157 (2012).

\bibitem{fqhe_mps1}
Michael P. Zaletel and Roger S. K. Mong, Phys. Rev. B {\bf 86}, 245305 (2012).

\bibitem{fqhe_mps2}
B. Estienne, Z. Papi\'c, N. Regnault, and B. A. Bernevig, Phys. Rev. B {\bf 87}, 161112 (2013). 

\bibitem{dmrg_zaletel} Michael P. Zaletel, Roger S. K. Mong, and Frank Pollmann, Phys.
Rev. Lett. {\bf 110}, 236801 (2013).

\bibitem{haldane_sphere} F. D. M. Haldane, Phys. Rev. Lett. {\bf 51}, 605 (1983).


\bibitem{rh85} F. D. M. Haldane and E. H. Rezayi, Phys. Rev. Lett. {\bf 54}, 237  (1985).


\bibitem{edge} Xin Wan, E. H. Rezayi and Kun Yang, Phys. Rev. B {\bf 68}, 125307 (2003).
\bibitem{zixiang_edge} Yuhui Zhang, Zi-Xiang Hu, and Kun Yang, Phys. Rev. B {\bf 88}, 205128 (2013).

\bibitem{yang} C. N. Yang, Rev. Mod. Phys. {\bf 34}, 694 (1962).


\bibitem{girvin_odlro}  S. M. Girvin and A. MacDonald, Phys. Rev. Lett. {\bf 58}, 1252 (1987).

\bibitem{read_odlro} N. Read, Phys. Rev. Lett. {\bf 61}, 1985 (1988).

\bibitem{rh88} E. H. Rezayi and F. D. M. Haldane, Phys. Rev. Lett. {\bf 61}, 17 (1988).

\bibitem{chak_odlro} Tapash Chakraborty and Wolfgang von der Linden, Phys. Rev. B {\bf 41}, 7872 (1990).

\bibitem{haldane_book} F. D. M. Haldane, in \emph{The Quantum Hall Effect}, R. E. Prange and S. M. Girvin, editors, New York: Springer, 1990.

\bibitem{rezayi_cylinder} E. H. Rezayi and F. D. M. Haldane, Phys. Rev. B {\bf 50}, 17199 (1994). 
\bibitem{soule} Paul Soul\'e and Thierry Jolicoeur, Phys. Rev. B {\bf 86}, 115214 (2012). 

\bibitem{tao-th1} R. Tao and D. J. Thouless, Phys. Rev. B {\bf 28}, 1142 (1983).
\bibitem{tao-th2} D. J. Thouless, Surf. Sci. {\bf 142}, 147 (1984).
\bibitem{jack} B. Andrei Bernevig and F. D. M. Haldane, Phys. Rev. Lett. {\bf 100}, 246802 (2008).


\end{thebibliography}
\end{document}